\documentclass[numberedappendix]{emulateapj}

\tightenlines

\countdef\decade=200
\advance\decade by \year
\countdef\hours=201
\advance\hours by \time
\divide\hours by 60
\countdef\mins=202
\advance\mins by \hours
\multiply\mins by 60
\multiply\hours by 100
\countdef\miltime=203
\advance\miltime by \hours
\advance\miltime by \time
\advance\miltime by -\mins

\newcommand{\gtsim}{\protect\raisebox{-0.5ex}{$\:\stackrel{\textstyle >}
        {\sim}\:$}}

\newcommand{\sigmad}{\sigma_{\rm d}}

\newcommand{\fmol}{f_{\rm H_2}}
\newcommand{\fatm}{f_{\rm HI}}
\newcommand{\ehat}{\hat{e}}

\newcommand{\calr}{\mathcal{R}}

\newcommand{\rmol}{r_{\rm H_2}}

\newcommand{\rd}{r_{\rm d}}

\newcommand{\zm}{z_{\rm H_2}}

\newcommand{\xd}{x_{\rm d}}
\newcommand{\xm}{x_{\rm H_2}}
\newcommand{\tauR}{\tau_{\rm R}}

\newcommand{\fdissoc}{f_{\rm diss}}

\newcommand{\mumol}{\mu_{\rm H_2}}
\newcommand{\taud}{\tau_{\rm d}}

\newcommand{\calf}{\mathcal{F}}

\newcommand{\ym}{y_{\rm H_2}}
\newcommand{\yd}{y_{\rm d}}

\newcommand{\xim}{\xi_{\rm H_2}}
\newcommand{\xid}{\xi_{\rm d}}
\newcommand{\Istar}{I^{*}}
\newcommand{\Jstar}{J^{*}}
\newcommand{\Estar}{E^{*}}
\newcommand{\Fstar}{F^{*}}
\newcommand{\bFstar}{\mathbf{F}^{*}}
\newcommand{\sigmae}{\sigma_{\rm e}}
\newcommand{\taueR}{\tau_{\rm eR}}
\newcommand{\taum}{\tau_{\rm HI}}
\newcommand{\nhi}{N_{\rm HI}}

\begin{document}

\title{The Atomic to Molecular Transition in Galaxies.\\I: An Analytic
Approximation for Photodissociation Fronts in Finite Clouds}

\slugcomment{Accepted to the Astrophysical Journal, August 8, 2008}

\author{Mark R. Krumholz\footnote{Hubble Fellow}}
\affil{Department of Astrophysical Sciences, Princeton University, Peyton Hall,
Princeton, NJ 08544 and Department of Astronomy \& Astrophysics, University of California, Santa Cruz, Integrative Sciences Building, Santa Cruz, CA 95060}
\email{krumholz@ucolick.org}

\author{Christopher F. McKee}
\affil{Departments of Physics and Astronomy, University of California, Berkeley, Campbell Hall,
Berkeley, CA 94720-7304}
\email{cmckee@astro.berkeley.edu}

\author{Jason Tumlinson}
\affil{Yale Center for Astronomy and Astrophysics, Yale University, PO Box 208121, New Haven, CT 06520 and Space Telescope Science Institute, 3700 San Martin Dr., Baltimore, MD 21218}
\email{tumlinson@stsci.edu}

\begin{abstract}
In this series of papers we study the structure of the atomic to molecular transition in
the giant atomic-molecular complexes that are the repositories of most molecular
gas in galaxies, with the ultimate goal of attaining a better understanding of what determines
galaxies' molecular content. Here we derive an approximate analytic solution for the
structure of a photodissociation region (PDR) in a cloud of finite
size that is bathed in an external dissociating radiation field. Our solution extends previous work, which
with few exceptions has been restricted to a one-dimensional treatment of the radiation field. We
show that our analytic results compare favorably to exact numerical
calculations in the one-dimensional limit. However, our more general geometry
provides a more realistic representation than a semi-infinite slab
for atomic-molecular complexes exposed to the interstellar radiation
field, particularly in environments such as low-metallicity dwarf
galaxies where the curvature and finite size of the atomic envelope
cannot be neglected. For clouds that are at least 20\% molecular we obtain analytic expressions for the molecular fraction in terms of properties of the gas and radiation field that are accurate to tens of percent, while for clouds of lower molecular content we obtain upper limits. As a side benefit, our analysis helps
clarify when self-shielding is the dominant process in H$_2$
formation, and under what circumstances shielding by dust makes a
significant contribution.
\end{abstract}


\section{Introduction}

In galaxies such as the Milky Way, where atomic and molecular phases
of the interstellar medium coexist, molecular clouds represent the
inner parts of atomic-molecular complexes \citep{elmegreen87}. The
bulk of the volume of
the interstellar medium is filled with far-ultraviolet (FUV) photons
capable of dissociating hydrogen molecules, and this radiation field
keeps the majority of the gas atomic. Gas that is predominantly
molecular is found only in dense regions where a combination of
shielding by dust grains and self-shielding by hydrogen molecules
excludes the interstellar FUV field. These molecular regions are
bounded by a photodissociation region (PDR) in which the gas is
predominantly atomic \citep[and references therein]{hollenbach99}.

To date most work on the structure of PDRs has been limited to one-dimensional geometries, including unidirectional or bidirectional beams of radiation impinging on semi-infinite slabs or purely radial radiation fields striking the surfaces of spheres \citep[e.g.][]{federman79, vandishoeck86, 
black87, sternberg88, elmegreen93, draine96,
hollenbach99, browning03, allen04}. For these one-dimensional problems the
literature contains both detailed numerical solutions and analytic
approximations for the problem of radiative transfer and
H$_2$ formation-dissociation equilibrium. These approaches yield good
results when the PDR is thin compared to the cloud as a whole, or for PDRs that are in close proximity to hot stars whose radiation and winds have compressed the PDR into a slab-like geometry. Many nearby well-studied PDRs, such as the Orion bar, fall into this latter category. However, the one-dimensional approximation is much less appropriate for giant clouds being dissociated by the combined starlight of many distinct stars and star clusters, particularly when the atomic region constitutes a significant fraction of the total cloud volume. The problem is especially severe in galaxies with low
metallicities and interstellar pressures, where the predominantly
molecular parts of cloud complexes generally constitute a small part
of the total mass and volume \citep[e.g.][]{blitz06b}. In this case
one cannot neglect either the curvature of the PDR or the finite
size of the molecular region, and a higher-dimensional approach is preferable. Previous work in one dimension is therefore of limited use for the problem on which we focus: determining the atomic and molecular content of galaxies on large scales, under the combined effects of all the sources of dissociating radiation in that galaxy.

Considerably less work has focused on higher-dimensional
geometries, since these require a treatment of the angular dependence of the radiation
field and its variation with position inside a cloud. As a result all
treatments of two- or three-dimensional radiation fields to date are purely numerical.
\citet{neufeld96} consider spherical clouds and
\citet{spaans97} allow arbitrary geometries, but their method
applies only in translucent clouds, and involves an approximate
numerical integration of the transfer equation. Similarly,
\citet{liszt00} and \citet{liszt02} present models for PDRs in
spherical clouds involving angular integration over the radiation
field in radial bins, coupled with a relaxation method to determine the
H$_2$ abundance at each radius. Neither of these approaches yield a
simple analytic estimate of the size of the PDR or the molecular
region, nor do they provide any insight into the dimensionless numbers
that can be used to characterize the problem of PDR structure. Such estimates, and the accompanying physical insights, would allow modeling of clouds over a wide range of galactic environments without the need for a complex and numerically costly radiative transfer calculation to cover each case.

Our goal in this work is to revisit the problem of determining the
size of a PDR in a finite cloud embedded in a multi-dimensional radiation field, and to derive analytic
approximations for the structure of a PDR that will yield gross yet observable
quantities such as the total atomic hydrogen column around a molecular
region and the fraction of a cloud's volume in the atomic and
molecular phases. As part of this work we determine the important
dimensionless numbers that characterize the problem, and we provide a rough
classification of PDRs based on them. The model we develop is capable
of spanning the range from galaxies where the gas in atomic-molecular
complexes is predominantly molecular and a slab treatment is
appropriate to dwarf galaxies where only a tiny fraction of the ISM
resides in the molecular phase. In future work (Krumholz, McKee, \&
Tumlinson, in preparation), we provide a more detailed application of
the results derived here to the problem of determining the atomic to
molecular ratios in galaxies. Before moving on, we do note that our focus on an analytic solution with a multi-dimensional radiation field, characterized by a few dimensionless numbers, has a price: our approach to the chemical and thermal physics of PDRs is significantly simpler than much previous work. We do not account for factors such as the temperature-dependence of rate coefficients or H$_2$ dissociation by cosmic rays. Our work is therefore less suited to making detailed predictions of the structures of individual PDRs than it is to making predictions for galactic-scale trends in atomic and molecular content.

We approach the problem of finite clouds by idealizing to the case of
a spherical cloud embedded in an isotropic radiation field, since this
allows us to explore the effects of finite cloud size and curvature
while at the same time keeping the problem simple enough to admit an
approximate analytic solution. Our approach is as follows: in
\S~\ref{formal} we state the formal problem and introduce some
physical approximations that are independent of geometry. In
\S~\ref{slab} we derive an approximate analytic solution to the
one-dimensional semi-infinite slab case, which allows us to
demonstrate the underlying physical principles of our approach. In
this section we also compare to a grid of numerical solutions and show
that our approach produces good agreement. Then in \S~\ref{sphere} we
extend our approach to handle the case of a spherical cloud embedded
in an isotropic radiation field. Finally we summarize and draw
conclusions in \S~\ref{conclusion}.

\section{The Formal Problem}
\label{formal}

Consider a region of hydrogen gas where the number density of hydrogen nuclei is $n$, mixed with
dust which has a cross section $\sigma_{{\rm d},\nu}$ per H nucleus
to radiation of frequency $\nu$. The hydrogen is a mix of atoms and
molecules, with a fraction $\fatm$ of the nuclei in the form of H\textsc{i} and
a fraction $\fmol=1-\fatm$ in the form of H$_2$. We consider
frequencies $\nu$ that fall within the Lyman-Werner (LW) band from
$\nu_1=c/1120\mbox{ \AA}$ to $\nu_2=c/912\mbox{ \AA}$, such that photons
of that frequency can be resonantly absorbed by hydrogen
molecules.

The equation of radiative transfer for a beam of radiation
in direction $\ehat$ passing through this gas is
\begin{equation}
\label{transferfreq}
\ehat \cdot \nabla I_{\nu} = -n \left(\frac{1}{2} \fmol \sigma_{{\rm
H}_2,\nu} + \sigma_{{\rm d},\nu}\right) I_{\nu},
\end{equation}
where $I_{\nu}$ is the radiation intensity at frequency $\nu$ and
$\sigma_{{\rm H}_2,\nu}$ is the cross section for absorption of
radiation at frequency $\nu$ by a molecule of hydrogen. The value of
$\sigma_{{\rm H}_2,\nu}$ may change with position as the fraction of
H$_2$ molecules in different quantum states changes. The
total fraction of the gas in the molecular phase is determined by the
balance between the rate of H$_2$ formation and dissociation,
\begin{equation}
\label{dissocfreq}
\fatm n^2 \calr
=
\frac{\fmol}{2} n \int d\Omega\, \int_{\nu_1}^{\nu_2} d\nu\,
\frac{I_{\nu}}{h \nu} \sigma_{{\rm H}_2,\nu} f_{{\rm diss},\nu}.
\end{equation}
where $\calr$ is the rate coefficient for formation of H$_2$ molecules
on dust grain surfaces and $f_{{\rm
diss},\nu}$ is the fraction of absorptions at frequency $\nu$ that
yield dissociation of the H$_2$ molecule rather than decay back to a
bound state.

Note that we do not include a source term in the transfer
equation (\ref{transferfreq}), because although most FUV photons
absorbed by H$_2$ molecules do decay through a vibrational ladder via
photon emission, the photons released in this process do not fall into
the LW band. Thus, the transfer equation we have written is only valid
for frequencies in the LW band. We have also neglected scattering of
FUV photons by dust grains. Since scattering is highly forward-peaked
at FUV wavelengths \citep[e.g.][]{roberge81}, this approximation is
reasonable as long as we take $\sigma_{{\rm d},\nu}$ to be the
absorption cross section, not the total cross section. We have also omitted other H$_2$ dissociation mechanisms other than LW photons, such as cosmic ray collisions and chemical reactions. These are significant only in nearly fully molecular regions where there are no significant numbers of LW photons present.

Equations (\ref{transferfreq}) and (\ref{dissocfreq}), together with
the atomic and dust physics that specify $\sigma_{{\rm H}_2,\nu}$ and
$\sigma_{{\rm d},\nu}$ and a boundary condition that specifies
$I_{\nu}$ on all rays entering the surface of a cloud, fully
determine $I_{\nu}$ and $\fmol$ at all positions. We cannot
solve them exactly, but we can obtain an approximation that exposes
the basic physical outlines of the solution. We begin by making two
standard approximations, following \citet{draine96}, to simplify the
atomic physics.

First, it is convenient to simplify the transfer equation
(\ref{transferfreq}) by dividing
by $h\nu$ to transform from intensity to photon number, and then by
integrating over frequency in the LW band. In so doing we can exploit
the fact that for realistic dust $\sigma_{{\rm d},\nu}$ is nearly
independent of frequency in the LW band \citep{draine96} to replace
$\sigma_{{\rm d},\nu}$ with a constant value $\sigmad$. Doing so gives
\begin{equation}
\label{transfer}
\ehat \cdot \nabla \Istar =  - n \sigmad \Istar -\frac{1}{2} n \fmol
\int_{\nu_1}^{\nu_2} d\nu\, \sigma_{{\rm H}_2,\nu} \Istar_{\nu},
\end{equation}
where $\Istar_{\nu}=I_{\nu}/(h\nu)$ is the photon number intensity,
i.e.\ the number of photons per unit
time per unit area per unit solid angle per unit frequency that cross
a given surface, and $\Istar=\int_{\nu_1}^{\nu_2} d\nu\, \Istar_{\nu}$ is
the photon number intensity integrated over the LW band.

Second, we note that $f_{{\rm diss},\nu}$ varies only weakly when
integrated over frequency and over position within a
PDR. \citet{draine96} show that over the width of a PDR it stays roughly
within the range $0.1-0.2$. Its value in free space depends on the assumed radiation spectrum; \citet{draine96} find $\fdissoc=0.12$ in free space for their fiducial choice, while \citet{browning03} suggest
$\fdissoc=0.11$ as a typical value. For simplicity we adopt a constant
value $f_{{\rm diss},\nu} = \fdissoc=0.1$ and
take this constant out of the integral, reducing the dissociation
equation to
\begin{equation}
\label{dissoc}
\fatm n^2 \calr = 
\frac{\fmol \fdissoc}{2} n \int d\Omega\, \int_{\nu_1}^{\nu_2} d\nu\,
\Istar_{\nu} \sigma_{{\rm H}_2,\nu}.
\end{equation}

It is convenient at this point to produce a combined
transfer-dissocation equation from (\ref{transfer}) and
(\ref{dissoc}). If we integrate equation (\ref{transfer}) over solid
angle $d\Omega$, we obtain
\begin{equation}
\nabla \cdot \bFstar = -n \sigmad c \Estar -
\frac{1}{2} \fmol n \int
d\Omega\, \int_{\nu_1}^{\nu_2} d\nu\, \sigma_{{\rm H}_2,\nu}
\Istar_{\nu},
\end{equation}
where
\begin{eqnarray}
\Estar & \equiv & \frac{1}{c} \int d\Omega \, \Istar \\
\bFstar & \equiv & \int d\Omega \, \ehat \Istar
\end{eqnarray}
are the photon number density and photon number flux integrated over
the LW band, respectively. We can then use equation (\ref{dissoc}) to
substitute for the last term, yielding the combined
transfer-dissociation equation
\begin{equation}
\nabla \cdot \bFstar = -n \sigmad c \Estar
- \frac{\fatm n^2 \calr}{\fdissoc}.
\label{transdissoc}
\end{equation}

\section{Solution in One Dimension}
\label{slab}

\subsection{Analytic Solution}
\label{slabanalyt}

We start by giving an approximate analytic solution to this formal
problem for unidirectional radiation impinging on a one-dimensional
semi-infinite slab in order to illustrate the physical principles
behind our approach. Consider a region of gas of density $n$ filling
the half-space $z>0$, subjected to a dissociating radiation field of
photon number intensity $\Istar = 4\pi \Jstar_0
\delta(|\ehat-\hat{z}|)$ that fills the half-space $z<0$, where $\Jstar_0$ is the angle-averaged intensity in free space. The corresponding free-space photon number density is $\Estar_0=4\pi \Jstar_0/c$, and the magnitude of the free-space photon flux is $\Fstar_0=c\Estar_0$. For simplicity we neglect the (relatively weak) temperature-dependence of $\calr$.

Since the
radiation intensity everywhere at all $z$ remains proportional to
$\delta(|\ehat-\hat{z}|)$, it immediately follows that
\begin{equation}
\bFstar = \Fstar \hat{z} = c \Estar \hat{z},
\end{equation}
at all points, and the combined transfer-dissociation equation reduces to
\begin{equation}
\label{slabeqn}
\frac{d\Fstar}{dz} = -n \sigmad \Fstar - \frac{\fatm n^2
\calr}{\fdissoc},
\end{equation}
subject to the boundary condition that $\Fstar = \Fstar_0$ at
$z=0$. Since numerical calculations show that the transition from
predominantly atomic gas to predominantly molecular gas in a PDR generally occurs in thin band bounded by much larger regions where the gas is either predominantly atomic or predominantly molecular, we can obtain a good approximation to the exact
solution by treating $\fatm$ as having a constant value near unity
over the bulk of the PDR, and then dropping to zero as a step function once the fully molecular surface is reached. For constant $\fatm=1$, we can
non-dimensionalize equation (\ref{slabeqn}) to
\begin{equation}
\label{slabnondim}
\frac{d\calf}{d\tau} = -\calf - \frac{1}{\chi},
\end{equation}
where $\calf=\Fstar/(\Fstar_0)$ is the fraction of the free-space
flux remaining, $\tau=n\sigmad z$ is the dust optical depth from the
slab surface, and 
\begin{equation}
\chi = \frac{\fdissoc \sigmad c \Estar_0}{n \calr}.
\end{equation}
Equation (\ref{slabnondim}) has the exact solution
\begin{equation}
\calf(\tau) = \frac{1}{\chi} \left[e^{-(\tau-\taum)}-1\right],
\end{equation}
where
\begin{equation}
\label{slabexact}
\taum = \ln\left(1+\chi\right)
\end{equation}
is the depth at which the flux goes to zero, which we take to be the optical depth through the H\textsc{i} region.
Of course in reality the flux should never go to zero exactly. That is
does in our solution is an artifact of our choice to treat $\fatm$
as constant. Nonetheless, since the transition from $\fatm\approx 1$ to
$\fatm \approx 0$ is sharp,
$\zm=\taum/(n\sigmad)$ should be a good approximation of the depth at
which the gas becomes predominantly molecular.

The dimensionless parameter $\chi/\fatm$
is the ratio of the two terms on the right hand side of equation
(\ref{slabeqn}) with $\Fstar$ set equal to its value $4\pi\Istar_0$ at
the slab edge. This makes its physical meaning clear: $\chi/\fatm$
represents the ratio of the absorption rate of LW photons by dust
grains to the absorption rate by H$_2$ molecules for a parcel of gas
exposed to the unattenuated free space radiation field. If the gas at
the edge of free space is predominantly atomic, as is the case for
example at the edge of an atomic-molecular complex, then $\fatm\approx 1$
and this ratio is simply given by $\chi$. For $\chi>1$ absorptions by
dust grains dominate, while for $\chi<1$ absorptions by H$_2$
molecules dominate.

For a giant molecular cloud in the Milky Way and its outer
atomic envelope, typical values of the number density, dust cross
section, and H$_2$ formation rate coefficient are $n\sim 30$
cm$^{-3}$, $\sigmad\sim 10^{-21}$ cm$^2$ and $\calr\sim
3\times 10^{-17}$ cm$^3$ s$^{-1}$ \citep{draine96}. Using the \citet{draine78} functional form for the local FUV radiation energy density as a function of wavelength, 
\begin{eqnarray}
\lambda E_{\lambda} & = & 6.84\times 10^{-14} \lambda_3^{-5}
\left(31.016\lambda_3^2-49.913\lambda_3+19.897\right)
\nonumber \\
& & \qquad \mbox{ erg cm}^{-3}
\end{eqnarray}
the free-space photon number density from $912-1120$ \AA\ is $\Estar_0=7.5\times 10^{-4}$
cm$^{-3}$. For an H$_2$ molecule in the ground state, this corresponds to a free-space dissociation rate of $3.24\times 10^{-11}$ s$^{-1}$. (In principle for a slab computation we should divide the observed value of $\Estar_0$ by two to account for the fact that one can only see half the sky at the surface of an opaque cloud, but we do not do so here because in \S~\ref{sphere} we will account for this effect self-consistently.) Thus, for Milky Way conditions not near a local strong source of FUV, $\chi$ of order a few might be typical. Thus, in the Milky Way dust shielding
is marginally significant in determining the structure of atomic-molecular complexes.

\subsection{The Two-Zone Approximation}
\label{slabapprox}

We can integrate the transfer-dissociation equation
(\ref{transdissoc}) directly in one dimension because, due to the
constant angular distribution of the radiation, there is a trivial
relationship between $\Estar$ and $\Fstar$. In multiple dimensions,
however, there is no simple relationship between the two, because the
angular distribution of the radiation intensity is not constant with
position inside a cloud. To overcome this problem, we adopt what we
call the \textit{two-zone approximation}. When the photon number density $\Estar$ is large, the
first term on the right hand side of equation (\ref{transdissoc}),
representing absorptions of photons by dust grains, is much larger
than the second term, representing absorptions by H$_2$
molecules. This makes intuitive sense: in regions where there many
dissociating photons present, the molecular fraction
will be very low, so there will be few H$_2$ molecules available to
absorb LW photons and most photons will be absorbed by dust. In regions
where $\Estar$ is small, the number of molecules will increase, and
for any appreciable number of molecules these will dominate the
absorption rate.

In the two-zone approximation, we divide the cloud
into a zone where dust absorption dominates and a zone where molecular
absorption dominates. In the dust-dominated region we drop the molecular absorption term in the radiative transfer or transfer-dissociation equations (equations \ref{transfer} or \ref{transdissoc}), and
approximate the opacity as having a constant value $n\sigmad$. In the zone where molecular absorptions dominate, we drop the dust absorption term in equation (\ref{transdissoc}) and approximate the molecular absorption term by $\phi n^2 \calr/\fdissoc$, where $\phi>1$ is a constant of order unity, whose precise value we determine below, that we include to account for the fact that some LW photons will be absorbed by dust grains even in the molecular-dominated region. We define a boundary between these two zones by the condition that the dust and molecular absorption  terms be equal, which is satisfied when
\begin{equation}
\label{dustcondition}
\frac{\Estar}{\Estar_0} = \frac{\phi}{\chi} \equiv \frac{1}{\psi},
\end{equation}
where we have set $\fatm=1$ because at the
point of equality the molecular fraction is $\ll 1$, and for convenience we have defined the modified dust to molecular absorption ratio $\psi = \chi/\phi$. With this approximation, the one-dimensional non-dimensionalized transfer-dissocation equation becomes
\begin{equation}
\label{slabtwozoneeq}
\frac{d\calf}{d\tau} = -\left\{
\begin{array}{lr}
\calf, & \calf > 1/\psi \\
1/\psi, & \calf < 1/\psi
\end{array}
\right..
\end{equation}

We shall see in \S~\ref{sphere} how the two-zone approximation enables
us to solve the problem in the spherical case. First, though, we examine the solution in the one-dimensional case. If $\psi<1$, then $\calf < 1/\psi$ is satisfied everywhere and equation (\ref{slabtwozoneeq}) has the trivial solution
\begin{equation}
\calf=\frac{\psi-\tau}{\psi}.
\end{equation}
The flux goes to zero at a depth $\taum = \psi$. If $\psi > 1$, the solution is
\begin{equation}
\calf = \left\{
\begin{array}{lr}
e^{-\tau}, & \tau < \taud \\
\psi^{-1} (\taum-\tau)/(\taum-\taud), & \tau > \taud
\end{array}
\right.,
\end{equation}
with
\begin{eqnarray}
\taud & = & \ln \psi \\
\taum & = & 1 + \ln \psi.
\end{eqnarray}
Here $\taud$ represents the dust depth into the slab at which the absorption
begins to be dominated by H$_2$ molecules, while $\taum$ is the
optical depth where we expect a transition from mostly atomic to
mostly molecular gas. Combining the two cases, we have
\begin{equation}
\label{slabtwozone}
\taum = \left\{
\begin{array}{lr}
\psi, & \psi < 1 \\
1+\ln\psi, & \psi > 1
\end{array}
\right..
\end{equation}

We now turn to the question of determining the constant $\phi$. Physically, we expect to have $\phi\rightarrow 1$ for $\chi \ll 1$, because in that case dust absorptions contribute negligibly throughout the cloud. For $\chi \gg 1$ we expect to have $\phi$ to asymptote to a value greater than unity, accounting for the contribution of dust to absorptions even in the molecular-dominated region. A comparison of the limiting behavior of the analytic solution (\ref{slabexact}) with the two-zone approximation (\ref{slabtwozone}) confirms this physical argument, and suggests that the appropriate limiting behavior is $\phi\rightarrow 1$ as $\chi\rightarrow 0$ and $\phi\rightarrow e$ as $\chi\rightarrow \infty$. We therefore adopt
\begin{equation}
\phi = \frac{2.5+\chi e}{2.5+\chi},
\end{equation}
which has the correct limiting behavior, and where the value $2.5$ is chosen to optimize agreement between the two-zone approximation and the analytic solution in the intermediate $\chi$ region.

\subsection{Comparison to Numerical Calculations}
\label{numeric}

Before using the two-zone approximation to compute the case of a
finite cloud with an isotropic radiation field, we check its accuracy
for the one-dimensional case by comparing with detailed numerical
calculations using the \citet{browning03} H$_2$ formation and
radiative transfer code. We refer readers to that paper for a full
description of the physics included in this calculation, but a brief
summary is that the code numerically integrates the
frequency-dependent equation of radiative transfer for a
unidirectional beam of radiation incident on an isothermal,
constant-density slab of gas mixed with dust. The transfer equation is
coupled to a statistical equilibrium calculation that determines the
populations of H\textsc{i} atoms and a large number of rotational and
vibrational levels of the H$_2$ molecule that are excited by LW band photons
in each computational
cell. The output of this calculation is the fraction of H nuclei in
molecules as a function of depth within the cloud. The code we use here differs from that described in \citet{browning03} only in that the earlier version accounted for absorptions of LW photons by dust grains by modifying the photodissociation rate using the method of \citet{vandishoeck86}, whereas the version we use here computes radiation attenuation by dust grains directly from the radiative transfer equation.

For the models we present here we use a density and temperature of
$n=5000$ cm$^{-3}$ and $T=90$ K. These values are chosen purely for
computational convenience, and have no significant impact on the
results. The incident radiation field is a unidirectional beam of photons uniformly distributed in frequency over the wavelength range $912-1120$ \AA. The frequency-dependent photon flux in this beam is $\Fstar_{\nu}$, so $\Estar_0=\Fstar_{\nu} (\nu_2-\nu_1)/c$ and $\Jstar_0=\Fstar_{\nu} (\nu_2-\nu_1)/(4\pi)$.
We adopt a dust extinction curve following
the functional form of \citet{cardelli89a}, scaled to give a dust cross
section per H nucleus at 1000 \AA\ of $\sigma=\sigma_{\rm d,MW} Z'$, where $Z'$
is the metallicity relative to solar and we take $\sigma_{\rm d,MW} = 6.0\times
10^{-22}$ cm$^2$ or $2.0\times 10^{-21}$ cm$^2$ to be two fiducial dust opacities for the Milky Way. These two values of $\sigma_{\rm d,MW}$ correspond to the estimated attenuation cross sections at 1000 \AA\ estimated by \citet{draine96} for dense and diffuse clouds Milky Way, respectively. We adopt a
rate coefficient for H$_2$ formation on grain surfaces of $\calr=\calr_{\rm MW} Z'$ with $\calr_{\rm MW}=3\times 10^{-17} Z'$ cm$^{-3}$ s$^{-1}$ as our fiducial Milky Way value \citep{wolfire08}. We do our computations for a grid of models running from $\Fstar_{\nu}=10^{-7}-10^{-3}$ photons
cm$^{-2}$ s$^{-1}$ Hz$^{-1}$
in steps of nine steps of 0.5 dex, from $Z'=10^{-2}-10^{0.5}$ in six steps of
0.5 dex, and for the two values of
$\sigma_{\rm d,MW}$ mentioned above. These values $\Fstar_{\nu}$ are significantly higher than are typical in the Milky Way, but are chosen so that, in conjunction with our choice of $n$, the ratio $\Estar_{\nu}/n$ that appears in $\chi$ is within the typical Milky Way range. With this parameterization
\begin{eqnarray}
\chi & = & 0.75 (\sigma_{\rm d,MW,-21}/\calr_{\rm MW}) (E_0^{*'} / n_2) \\
& = & 4.07 \sigma_{\rm d,MW,-21} \Fstar_{\nu,-5}
\end{eqnarray}
where $\sigma_{\rm d,MW,-21}=\sigma_{\rm d,MW}/10^{-21}$ cm$^2$ is the Milky Way 1000 \AA\ dust absorption opacity normalized to $10^{-21}$ cm$^2$, $E_0^{*'}=E_0^*/7.5\times 10^{-4}$ cm$^{-3}$ is the free-space dissociating photon number density normalized to the Milky Way value, $n_2 = n/100$ cm$^{-3}$ is the number density of hydrogen nuclei in units of 100 cm$^{-3}$, and 
$\Fstar_{\nu,-5}=\Fstar_{\nu}/10^{-5}$ photons cm$^{-2}$ s$^{-1}$ Hz$^{-1}$.
Thus the
calculation covers a broad range of parameters from strongly
dust-dominated to strongly molecular-dominated, thereby bracketing the real Milky Way value of $\chi\sim 1$. Note that $\chi$ is
independent of $Z'$ because for the parameterization we have chosen the $Z'$-dependences of $\sigmad$ and $\calr$ cancel. Since we predict that the dust optical depth through
the PDR, $\taum = n\sigmad \zm$, depends only on $\chi$, and $\chi$ in turn depends only on the ratio $\sigmad/\calr$, $\taum$ should be independent of $Z$. Since
we use a range of $10^{2.5}$ in $Z'$, our numerical calculations represent a
strong test of this prediction.

\begin{figure}
\plotone{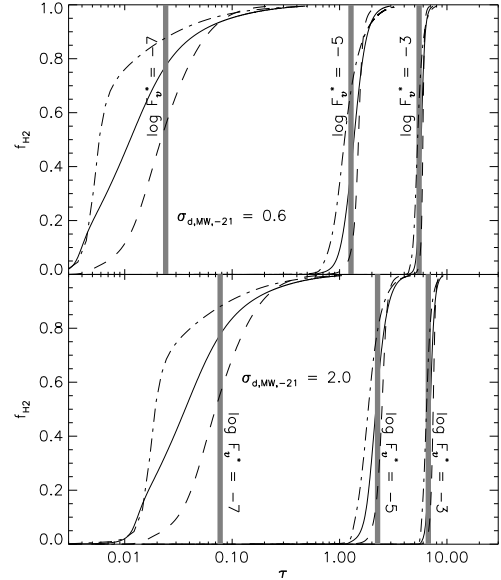}
\caption{
\label{fh2prof}
The plots show $\fmol$ versus dust optical depth $\tau=n\sigmad z$ for
our numerical radiative transfer calculations with $\log Z'=-1.5$
(\textit{dashed lines}), $\log Z'=-0.5$ (\textit{solid lines}), and
$\log Z'=0.5$ (\textit{dot-dashed lines}). The gray vertical lines
indicate the optical depth of the transition to fully molecular as
calculated with the two-zone approximation, equation
(\ref{slabtwozone}). Each cluster of three
curves plus a vertical line indicating a prediction
corresponds to a radiation flux $\log F_{\nu}^*=-7$, $-5$, or
$-3$, as indicated. The two panels are for the cases
$\sigma_{\rm d,MW,-21}=0.6$ and $2.0$, as indicated.
}
\end{figure}

\begin{figure}
\plotone{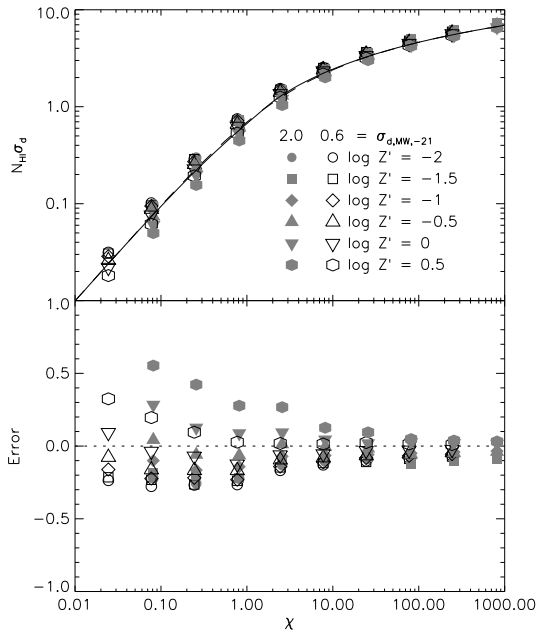}
\caption{
\label{hicolcomp}
The upper panel shows the dust opacity $\nhi\sigmad$
through the numerically-determined H\textsc{i} column (\textit{various
symbols}) as a function of $\chi$. The values of $Z'$ and
$\sigma_{\rm d,MW,{-21}}$ for each calculation are indicated by the plot
symbol. This is compared to the optical depth $\taum$ computed from
the two-zone approximation (equation \ref{slabtwozone}, \textit{thin solid
line}) and computed using the analytic solution (equation \ref{slabexact},
\textit{thick dashed line}). The lower panel shows the error in the two-zone
approximation, defined as $\mbox{Error} = \nhi\sigmad/\taum-1$; the dotted line indicates zero error.
}
\end{figure}

Figure \ref{fh2prof} shows $\fmol$ versus depth within a cloud as
computed numerically for a sample of our input parameters, overlaid
with the corresponding locations of the atomic to molecular transition
as calculated via the two-zone approximation. As the figure shows, the
two-zone approximation does a very good job reproducing the location
of this transition over an extremely broad range of parameters.

To quantify the quality of the approximation, we must define a fiducial measure for the depth of the H$_2$ region in the numerical calculations, since in these runs $\fatm$ approaches but never reaches unity. The most reasonable measure is
\begin{equation}
\label{nhiintegral}
\nhi = \int_0^{\infty} dz\, \fatm n,
\end{equation}
the total H\textsc{i} column integrated through the cloud. Since the radiation field is attenuated exponentially or faster, and  $\fatm$ is proportional to radiation intensity in the region where $\fmol\approx 1$, this integral is guaranteed to converge. In the limit where the transition from HI to H$_2$ is sharp it approaches the total gas column up to the transition point. In practice we cannot continue the numerical integration to $z=\infty$, so we truncate the integral at the value of $z$ where $\fatm=5\times 10^{-3}$; using $\fatm=10^{-2}$ instead changes the value by less than 8\% for all our runs, and by less than 2\% for all runs with $\chi>0.1$, so our evaluation of the integral should be accurate to this level.

We plot the dust opacity through this hydrogen column, $\nhi\sigmad$, and the corresponding value $\taum$ predicted by
the two-zone approximation, in Figure \ref{hicolcomp}. As the figure shows, the two-zone
approximation recovers the numerically-computed H\textsc{i} column to
better than 50\% accuracy over almost a five-decade range in
$\chi$. The error in the two-zone approximation is generally
comparable to or smaller than the spread between models with different
dust opacities but the same value of $\chi$. 

The error in our approximation is largest at small $\chi$, and examination of Figure \ref{fh2prof} suggests the reason why: by evaluating the equations with $\fatm=1$ inside the PDR, we have assumed that the transition from atomic to molecular is sharp. This is true for $\chi\sim 1$ or greater, but begins to fail for $\chi \ll 1$. In our runs with $\chi\ge 1$, typically 95\% of the gas is atomic in the region where $\fatm>0.5$; and even at the depth where $\fatm$ drops to $5\times 10^{-3}$ more than half the gas column above that point is atomic. This indicates a very sharp atomic-molecular transition, so our approximation that $\fatm=1$ until the gas is almost entirely molecular is a good one. For $\chi \approx 0.01$, on the other hand, roughly 80\% of the gas at $\fatm<0.5$ is atomic, and H\textsc{i} contributes only 10\% of the total gas column above $\fatm=5\times 10^{-3}$. The transition from atomic to molecular is therefore much more gradual, and our accuracy suffers as a result.
 
Nonetheless, we note that $\chi \ll 1$ does not appear to be physically realized in normal galactic environments. For Milky Way molecular clouds $\chi\sim 1$ or greater, and reducing $\chi$ to 0.01 would require some combination of reducing the ISRF and increasing the atomic gas density by a factor of 100. Such a combination of very high atomic ISM density and very low radiation field is generally not observed. We conclude that, for realistic physical parameters, and given that these parameters (such as
$\sigma_0$ and $\calr$) are themselves uncertain at the factor of a
few level \citep[e.g.][]{wolfire08}, the error in the two-zone
approximation is unlikely to be the dominant one.

\section{Solution for Spherical Clouds}
\label{sphere}

\begin{figure}
\epsscale{0.7}
\plotone{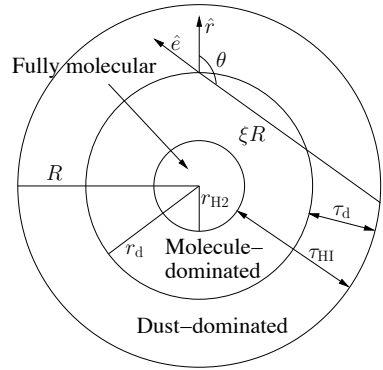}
\epsscale{1.0}
\caption{
\label{diagram}
Illustration of the two-zone approximation in spherical geometry.\\\\
}
\end{figure}

We now extend the two-zone approximation to a spherical cloud of
radius $R$ embedded in a uniform, isotropic radiation field of
angle-averaged intensity $\Jstar_0$. (Note that this radiation field has the same LW
photon number density as the unidirectional radiation field considered
in \S~\ref{slab}, so it gives the same dissociation rate in free
space.) Figure \ref{diagram} illustrates the
basic geometry of the problem and our approximation: we consider the
dust-dominated region to run from radius $r=\rd$ to $r=R$, and the
molecular self-shielding region to run from $r=\rmol$ to $r=\rd$. For
convenience we introduce the dimensionless position variables $x=r/R$
and $y=1-x$, and we define the dust optical depths from the surface to $\rd$ and to $\rmol$ as $\taud$ and $\taum$, respectively.

In \S~\ref{spheredust} and \S~\ref{spherenodust}, we develop the basic
equations that describe the two-zone approximation for clouds with and
without dust opacity-dominated envelopes. We then explore three
limiting cases of these equations. We consider the behavior at the
boundary between the presence and absence of a dust-dominated
zone in \S~\ref{dustboundary}, and we explore several interesting limits in \S~\ref{limitsection}.
We then give a
numerical solution and an analytic approximation to it in
\S~\ref{numericsol}. In \S~\ref{planarcomp} we compare our solution
for a finite cloud to the standard slab approximation, to determine
when the slab approximation is valid and when it fails. In \S~\ref{error},
we address the level of uncertainty introduced by the approximations we 
make in the spherical case. Finally,
in \S~\ref{example} we present some example calculations using our
analytic approximation.

\subsection{Clouds with Dust-Dominated Zones}
\label{spheredust}

First consider the case where $\psi$ is large enough so that there
is a dust-dominated zone in the outer part of the cloud where
molecular self-shielding is negligible, i.e.\ $\rd < R$. The transfer
equation in this region becomes
\begin{equation}
\label{transferdust}
\ehat \cdot \nabla \Istar =  - n \sigmad \Istar,
\end{equation}
which for rays originating at the cloud surface and staying entirely
within the dust-dominated region has the trivial solution
\begin{equation}
\Istar(x, \mu) = \exp(-\tauR\xi) \Jstar_0,
\end{equation}
where $\tauR=n\sigmad R$ is the center-to-edge dust optical depth of
the cloud,
\begin{equation}
\label{xidef}
\xi = \left(\sqrt{1 - x^2 + x^2\mu^2} - x\mu\right),
\end{equation}
is the distance, normalized to the cloud radius, from radius $r$ to
the cloud surface on a ray that makes an angle $\theta$ relative to
the radial vector (see Figure \ref{diagram}), and
$\mu=\cos\theta=-\ehat\cdot\hat{r}$. This solution applies for $\mu >
0$. On the other hand, if $\mu<\mumol$, 
\begin{equation}
\label{mumoleqn}
\mumol \equiv -\sqrt{1-\left(\frac{\xm}{\xd}\right)^2},
\end{equation}
then the ray passes through a part of the cloud that is fully
molecular. Since the fully molecular region will be extremely opaque,
to good approximation along these rays $\Istar(x,\mu)=0$.
Finally, rays for which $0 > \mu > \mumol$ pass through the region
where the opacity is dominated by molecules rather than dust,
but where the gas is not yet fully molecular. Again, we use the
approximation that the transition from a low molecular
fraction to fully molecular is sharp, so that over most of this region
the molecular fraction is not vastly larger than it is at the region's
edge. This is consistent with numerical solutions of the problem (e.g.\ Figure \ref{fh2prof}),
which show that the atomic fraction $\fatm$ is nearly constant through
the bulk of the PDR, and rises from its free space value to
unity over a small region. This means that the opacity is not much
greater than its value
of $n\sigmad$ at the outer edge of the molecule-dominated region. We
therefore approximate that the optical depth along these rays is the
same as for those in the dust-dominated region, $\tauR\xi$. This approximation is perhaps the least certain part of our calculation, and we quantify the level of uncertainty that it produces in \S~\ref{error}.
Combining these three regions, we have an
approximate intensity
\begin{equation}
\label{istarray}
\Istar(x, \mu) = \left\{
\begin{array}{lr}
\exp(-\tauR\xi) \Jstar_0, & \mu > \mumol \\
0, & \mu < \mumol
\end{array}
\right..
\end{equation}

The location of $\xd$, the crossover point from dust- to
molecule-dominated absorption, is defined by the condition that
$\Estar/\Estar_0=1/\psi$ (equation \ref{dustcondition}). For
convenience we define the function $\eta_0$ by
\begin{equation}
\label{f0def}
\frac{\Estar(\xd)}{\Estar_0} = \eta_0(\xd, \xm; \tauR) = 
\frac{1}{2} \int_{\mumol}^1 d\mu\, \exp\left(-\tauR \xid \right),
\end{equation}
where $\xid=\xi(\xd)$.
We show that $\eta_0$ can be evaluated in terms of exponential
integrals in Appendix \ref{fintegrate}. The first equation for the
two-zone approximation is therefore
\begin{equation}
\label{f0eqn}
\eta_0(\xd,\xm;\tauR) = \frac{1}{\psi}.
\end{equation}

Inside $\xd$, we drop the dust opacity term, so that the combined
transfer-dissociation equation (\ref{transdissoc}) becomes
\begin{equation}
\frac{1}{r^2}\frac{d}{dr}(r^2\Fstar) = \frac{\phi n^2 \calr}{\fdissoc},
\end{equation}
where we have again assumed that $\fatm\approx 1$ outside the fully
molecular region, and for convenience we have inverted the sign by
defining $\bFstar=-\Fstar \hat{r}$. The solution is
\begin{equation}
\label{fluxsol}
\Fstar = \frac{\tauR}{3\psi} x
\left[1-\left(\frac{\xm}{x}\right)^3\right] \Fstar_0,
\end{equation}
where we have chosen the constant of integration by requiring that
$\Fstar=0$ at $x=\xm$. To determine $\xm$ from the boundary
conditions, however, we must determine the flux at some other
location. Thus we evaluate the flux $\Fstar$ at $\xd$ by integrating
the intensity over solid angle using equation (\ref{istarray}). For
convenience we define the function $\eta_1$ by
\begin{equation}
\label{f1def}
\frac{\Fstar(\xd)}{\Fstar_0} = \eta_1(\xd, \xm; \tauR) 
=\frac{1}{2} \int_{\mumol}^1 d\mu\,
\mu \exp\left(-\tauR \xid \right).
\end{equation}
As with $\eta_0$, in Appendix \ref{fintegrate} we evaluate $\eta_1$ in terms
of exponential integrals. Combining (\ref{f1def}) with (\ref{fluxsol})
gives an implicit equation for $\xm$:
\begin{equation}
\eta_1(\xd,\xm; \tauR) = 
\frac{\tauR \xd}{3 \psi} \left[1-\left(\frac{\xm}{\xd}\right)^3\right].
\label{f1eqn}
\end{equation}
Together, equations (\ref{f0eqn}) and (\ref{f1eqn}) constitute two
equations in the two unknowns $\xd$ and $\xm$, and thus fully
determine the location of the transition from predominantly atomic to
predominantly molecular in the two-zone approximation.

\subsection{Clouds without Dust-Dominated Zones}
\label{spherenodust}

Now consider the case where $\psi$ is small enough so that
there is only one zone, because even gas at the edge of the cloud is
sufficiently molecular for absorptions by molecules to outnumber those
by dust grains. In this case equation (\ref{fluxsol}) applies
throughout the cloud, so we must fix $\xm$ directly from the boundary
conditions. To do so we need to know the flux
$\Fstar(1)$ at the cloud surface. This is not simply
$\Fstar_0=c\Estar_0$ as in the case of a unidirectional radiation field; in free space for an isotropic radiation field $\Fstar$ vanishes, and $\Fstar(1)$ is non-zero only because rays
passing through the cloud do not carry the same intensity as rays
that do not pass through it, preventing the integral over angle from
vanishing. Thus for a sufficiently transparent cloud, $\Fstar(1)$ approaches zero, its value in free space. Conversely, at the surface of a cloud that is opaque and extremely large, $\Fstar(1)=\Fstar_0/4$. The factor of $1/4$ relative to the unidirectional case arises because because half the solid angle is blocked by an opaque object (providing one factor of $1/2$), and because in the part of the sky that is not blocked the radiation is isotropic, and one must average over all the directions in which photons are traveling to find the fraction of that motion in the $-\hat{r}$ direction (providing another factor of $1/2$).

The problem of determining the intensity is exactly the same as in
\S~\ref{spheredust}. At the surface of the cloud rays at
angles $\mu>0$ do not pass through the cloud and therefore contribute
the unattenuated free space intensity $\Istar_0$. Those with $\mu <
\mumol = -(1-\xm^2)^{1/2}$ pass through the fully molecular region and
therefore contribute zero intensity. For rays at angles $\mumol < \mu
< 0$, we make the same approximation as in \S~\ref{spheredust}, that
the molecular absorption rate per unit distance that a photon travels is
roughly constant until one approaches the sharp transition from atomic
to molecular. Thus the molecular opacity-dominated part of the PDR has
a constant effective opacity, which we can determine by computing its
value at the cloud surface. For convenience we characterize this
opacity via an effective cross-section per H nucleus $\sigmae$. By
examining the transfer-dissociation equation (\ref{transdissoc}), it
is clear that for $\fatm\approx 1$ this opacity is
\begin{equation}
\label{sigmaedef}
\sigmae = \frac{\phi n \calr}{\fdissoc c \Estar(1)}.
\end{equation}

With this approximation the transfer equation through the region
outside where the gas becomes fully molecular is simply equation
(\ref{transferdust}) with $\sigmad$ replaced by $\sigmae$, and the
solution for the intensity along each ray is given by equation
(\ref{istarray}) with $\tauR$ replaced by $\taueR =
(\sigmae/\sigmad)\tauR$. The photon number density and flux at the
cloud surface are therefore given by
\begin{eqnarray}
\frac{\Estar(1)}{\Estar_0} & = & \eta_0(1, \xm; \taueR) \\
\frac{\Fstar(1)}{\Fstar_0} & = & \eta_1(1, \xm; \taueR).
\end{eqnarray}
With these arguments, evaluating equations (\ref{f0eval}) and
(\ref{f1eval}) shows that $\eta_0$ and $\eta_1$ reduce to
\begin{eqnarray}
\label{f0xd1}
\eta_0(1,\xm;\taueR) & = & \frac{1-e^{2\mumol\taueR}}{4\taueR}+\frac{1}{2} \\
\label{f1xd1}
\eta_1(1,\xm;\taueR) & = &
\frac{(1-2\mumol\taueR)e^{2\mumol\taueR}-1}{8\taueR^2} + \frac{1}{4}.
\end{eqnarray}
Using these values of $\Estar(1)$ and $\Fstar(1)$ in equations
(\ref{sigmaedef}) and (\ref{fluxsol}), we find
\begin{eqnarray}
\label{f0eqnodust}
\frac{1-e^{2\mumol\taueR}}{4\taueR}+\frac{1}{2} &
= & \frac{\tauR}{\psi \taueR} \\
\label{f1eqnodust}
\frac{(1-2\mumol\taueR)e^{2\mumol\taueR}-1}{8\taueR^2} + \frac{1}{4}
& = & \frac{\tauR}{3\psi} \left(1-\xm^3\right).
\end{eqnarray}
We therefore again have two equations in two unknowns, with the
unknowns in this case being $\taueR$ and $\xm$. Note that in this case
$\mumol$ can be isolated in the first equation, to give
\begin{equation}
\label{musol}
\mumol = \frac{1}{2\taueR}
\ln\left(1+2\taueR-\frac{4\tauR}{\psi}\right).
\end{equation}
Together with the relation between $\mumol$ and $\xm$ (equation \ref{mumoleqn}), this reduces the problem to a single non-linear equation, which is
convenient for numerical solution. However, the two-equation form is
more convenient for an analytic approach.

At this point it is worth making a few remarks about the behavior of equations (\ref{f0eqnodust}) and (\ref{f1eqnodust}). First, equation (\ref{f0eqnodust}) implies that $\tauR/(\psi\taueR)>1/2$, so the argument of the logarithm in equation (\ref{musol}) is always less than unity and $\mumol$ is negative. Second, in all of these equations $\tauR$ and $\psi$ appear only through the combination $\tauR/\psi$, so values of $\xm$ and $\taueR$ must be constant on lines of constant $\tauR/\psi$. Finally, note that in \S~\ref{numericsol} we give an approximate analytic solution to equations (\ref{f0eqnodust}) and (\ref{f1eqnodust}).

\subsection{The Dust-Dominated Zone Boundary}
\label{dustboundary}

\begin{figure}
\plotone{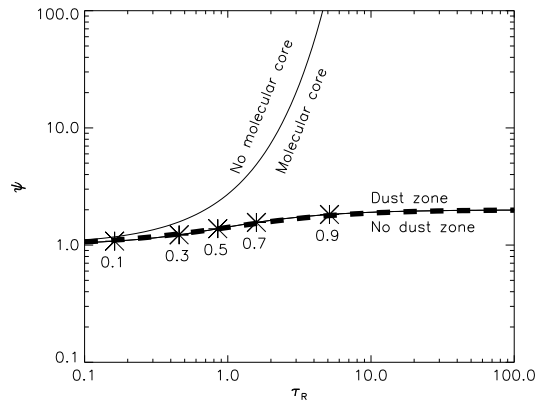}
\caption{
\label{chitauzone}
The thin black curves show the boundaries in the $(\tauR,\psi)$-plane at
which $\xd=1$, $\taueR=\tauR$, and at which $\xm=0$, as indicated by
the text accompanying each curve. The thick dashed curve shows our
approximation to the $\xd=1$ curve, equation (\ref{weakapprox}).
The asterisks along the curve for $\xd=1$ mark the points at which
$\xm=0.1$, $0.3$, $0.5$, $0.7$, and $0.9$, as indicated.
}
\end{figure}

We first identify the boundary between the presence and absence
of a dust-dominated region. In the case of a perfectly beamed
radiation field impinging on a semi-infinite planar slab, which we treated
in \S~\ref{slab}, this is $\psi=1$. The result is more complex in the case of
an isotropic radiation field and a cloud of finite size. The boundary
between the two cases is defined by the condition that $\xd=1$ or
$\taueR=\tauR$, i.e.\ that the cross-over between dust-dominated and
molecular-dominated absorptions occur at the cloud surface, or
equivalently that the dust and molecular effective opacities at the
cloud surface are equal.
It is immediately obvious that equations (\ref{f0eqn})
and (\ref{f1eqn}) become identical to equations (\ref{f0eqnodust}) and
(\ref{f1eqnodust}) in this limit. If we set $\taueR=\tauR$, then
equations (\ref{musol}) and (\ref{f1eqnodust}) define a curve in the
$(\tauR,\psi)$-plane that
corresponds to the point where dust-dominated layer disappears. 
Above the curve the radiation field is
strong enough so that the outer part of the cloud is dust
opacity-dominated, while below it molecular opacity dominates
throughout. Along the bounding curve, 
\begin{equation}
\xm = \sqrt{1 - \frac{1}{4\tauR^2}
\left[\ln\left(1+2\tauR-\frac{4\tauR}{\psi}\right)\right]^2}.
\end{equation}
We show the curve at which the dust-dominated layer vanishes, and the
value of $\xm$ along the curve, in Figure \ref{chitauzone}.

That along this curve $\psi\rightarrow 1$ for $\tauR\ll 1$ and
$\psi\rightarrow 2$ for $\tauR\gg 1$ makes intuitive sense. If
$\tauR\ll 1$, the dust optical depth through the cloud is tiny and so
the radiation field has its unattenuated, free-space value regardless
of position in the cloud. Thus the condition that dust shielding and
molecular shielding contribute equally at the cloud surface
($\xd= 1$) can only be fulfilled is they are equal or nearly
so in free space, which is simply a requirement that
$\psi=1$. Similarly, if $\tauR \gg 1$ then the cloud is
effectively a semi-infinite slab whose curvature is negligible. In
this case the radiation field at the cloud surface only contains
contributions from rays with $\mu>0$, i.e. those that never pass
through the cloud; rays with $\mu<0$ are infinitely attenuated. Thus,
the radiation field has exactly half its free-space value, the
molecular fraction has double its free-space value, and the
condition that dust shielding and molecular shielding are equal
reduces to the requirement that $\psi=2$.

These considerations suggest that a function that interpolates between
these two limiting behaviors is likely to produce a good
approximation. Numerical experimentation shows that the curve
\begin{equation}
\label{weakapprox}
\psi \approx \frac{1.4+2\tauR}{1.4+\tauR}
\end{equation}
reproduces the true value of $\psi$ along the curve $\xd=1$,
$\taueR=\tauR$ to better than 3\% for all $\tauR$. We show this
approximate solution with the thick dashed curve in Figure \ref{chitauzone}.

\subsection{Limiting Cases}
\label{limitsection}

We can better understand the behavior of PDRs in finite clouds by exploring several limiting cases of our equations, corresponding to clouds that are very large or very small, and to radiation fields that are very strong or very weak.

\textit{Case 1: Strong Radiation Fields.} The first limit we consider is one in which the radiation field is so strong that there is no fully molecular core, so $\xm=0$. It is easy to verify that if there is no dust-dominated region, so equations (\ref{f0eqnodust}) and (\ref{f1eqnodust}) apply, then there are no finite values of $\psi$ and $\tauR$ such that $\xm=0$. (However, see \S~\ref{error}, where we show that this behavior is probably not physical.) On the other hand, if there is a dust-dominated region, equations (\ref{f0eqn}) and (\ref{f1eqn}) apply and $\xm=0$ can be reached at finite $\psi$ and $\tauR$. This becomes clear if we note that $\xm=0$ implies $\mumol=-1$ (following equation \ref{mumoleqn}), and equation (\ref{f1eqn}) then admits the solution $\xd=\eta_1(\xd,0;\tauR)=0$. Since $\xd=0$, it immediately follows that $\xi_{\rm d}=1$ and $\eta_0(0,0;\tauR)=e^{-\tauR}$. Equation (\ref{f0eqn}) then gives $\psi=e^{\tauR}$. The physical meaning of this solution is that $\psi=e^{\tauR}$ is the critical curve along which $\xm=\xd=0$; at this value of $\psi$ or larger the radiation field is too intense for a fully molecular core to exist. We plot the critical curve in Figure \ref{chitauzone}. In Appendix \ref{strongappendix} we solve
equations (\ref{f0eqn}) and (\ref{f1eqn}) perturbatively in the vicinity of the critical curve, and show that
near the strong radiation boundary the
solution is
\begin{eqnarray}
\xd & = & \left[\frac{6}{\tauR (\tauR+2)} \left(\frac{e^{\tauR}}{\psi}-1\right)\right]^{1/2} \\
\xm & = & \left[\frac{1536}{25\tauR(\tauR+2)^3}\right]^{1/4} \left(\frac{e^{\tauR}}{\psi}-1\right)^{5/4}.
\end{eqnarray}
This solution obviously only applies for $e^{\tauR} \geq \psi$.

\textit{Case 2: Small Clouds.} Before analyzing this case, we warn that in \S~\ref{error} we show that our solution in this case should be regarded as giving an upper limit on the molecular fraction rather than a direct estimate. However it is still useful to consider this case, both in order to derive upper limits and to provide expressions that can be incorporated into approximations in parts of parameter space where our method does provide estimates rather than upper limits. We have shown that for $\psi>1$, there is a finite value of $\tauR$ at which the fully molecular core vanishes, and conversely that if $\psi<1$ there is no finite $\tauR$ for which $\xm=0$. However, one can easily verify that when there is no dust-dominated zone ($\psi<1$) and
equations (\ref{f0eqnodust}) and (\ref{f1eqnodust}) apply, 
$\xm\rightarrow 0$ as $\tauR\rightarrow 0$. Physically, this corresponds to the case of a cloud that is so small that its dust is optically thin to LW photons. In Appendix
\ref{largesmallappendix} we show that in this limit the solution may be
approximated by
\begin{eqnarray}
\label{smallcloudeqn1}
\taueR & = & \frac{\tauR}{\psi} + \frac{\tauR^2}{2\psi^2} +
\frac{\tauR^3}{4\psi^3} + \frac{7\tauR^4}{60\psi^4} \\
\xm & = & \frac{\tauR}{\sqrt{3}\psi} + \frac{2\sqrt{3}\tauR^2}{5\psi^2}.
\end{eqnarray}
Note that by definition $\taueR/\tauR = [\Estar_0/\Estar(1)]/\psi$, so equation (\ref{smallcloudeqn1}) is effectively a series expansion for the photon number density at the cloud surface: $\Estar(1)=\Estar_0 [1 - \tauR^2/(2\psi) + \cdots]$. Thus the leading-order approximation reduces to the statement that a small cloud blocks no radiation in any direction, so $\Estar(1)=\Estar_0$, the unattenuated value. The next-order correction accounts for the small fraction of photons that are blocked at the cloud surface.

\textit{Case 3: Large Clouds.} Our final limiting case is that of a cloud so large that
the transition from atomic to molecular gas occurs in a thin layer at
the cloud surface, so that the cloud's curvature is negligible. Before
proceeding we note that this case is \textit{not} the same as the
case of a one-dimensional slab subject to a unidirectional beam of
radiation that we analyzed in \S~\ref{slab}. The difference is that
here the radiation field is isotropic, so it has an angular dependence
that can vary with depth within the cloud. For this reason the large cloud limit with an isotropic radiation field is a
two-dimensional problem even if the cloud is a semi-infinite slab.
To analyze this case we perform a series expansion around the limit
$\tauR\rightarrow \infty$, but with $\ym\tauR$ finite, so that there
is a finite optical depth to the molecular region. If $\psi>2$ then a
dust-dominated zone exists, and we solve this problem by starting from
equations (\ref{f0eval}) and (\ref{f1eval}) and series expanding $\eta_0$
and $\eta_1$ to first order in $\tauR^{-1}$. Doing so gives
\begin{eqnarray}
\eta_0(\xd,\xm;\tauR) & = & \frac{{\rm E}_2(\taud)}{2} + \frac{e^{-\taud}+\taud {\rm
E}_2(\taud)}{4\tauR} \\
\eta_1(\xd,\xm;\tauR) & = & \frac{{\rm E}_3(\taud)}{2} + 
\frac{\taud{\rm E}_3(\taud)}{2\tauR},
\end{eqnarray}
where $\taud=\yd\tauR$. Equation (\ref{f0eqn}) therefore becomes
\begin{equation}
\label{taupeqn}
\frac{1}{\psi} = \frac{{\rm E}_2(\taud)}{2} + \frac{e^{-\taud}+\taud {\rm
E}_2(\taud)}{4\tauR}
\end{equation}
to first order in $\tauR^{-1}$, which is straightforward to solve
numerically to determine $\taud$ for a given $\psi$ and $\tauR$. Alternately, one may obtain a purely analytic expression by dropping the $1/\tauR$ correction term. In this case the equation becomes ${\rm E}_2(\taud) = 2/\psi$, which has the approximate solution
\begin{equation}
\taud \approx 0.83 \, \ln(0.2\psi + 0.6);
\end{equation}
this expression is accurate to better than 2\% for $2 \le \psi \le 100$. Since
$\xd=1-\taud/\tauR$, this fixes $\xd$. Similarly, once
$\taud$ is known it is straghtforward to solve equation
(\ref{f1eqn}) to first order in $\tauR^{-1}$ to obtain
\begin{equation}
\xm = 1 - \left[\taud+\frac{\psi}{2} {\rm E}_3(\taud)\right]
\tauR^{-1}.
\end{equation}

\begin{figure}
\plotone{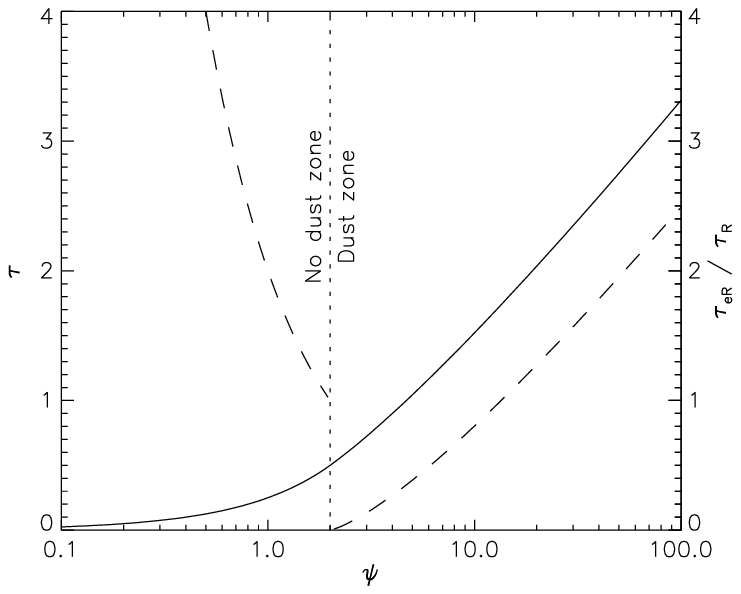}
\caption{
\label{planeplot}
The plot shows the solution in the large cloud limit. The curves shown
are the dust optical depth to the fully molecular
region $\taum=\tauR(1-\xm)$ (\textit{solid curve}), the dust optical depth to the
point of dust-molecular absorption equality $\taud=\tauR(1-\xd)$ (\textit{dashed
curve to the right of $\psi=2$}), and the ratio of the effective molecular
opacity to the dust opacity $\taueR/\tauR$ (\textit{dashed
curve to the left of $\psi=2$}). The dotted vertical line at $\psi=2$
indicates the boundary between the presence and absence of a
dust-dominated zone in the weak radiation limit.
}
\end{figure}

For $\psi<2$, there is no dust-dominated zone, and we must instead
solve equations (\ref{f0eqnodust}) and (\ref{f1eqnodust})
in the limit $\tauR\rightarrow \infty$. We note that for a very large cloud $\Estar(1)=\Estar_0/2$ because the cloud blocks half the sky, and it therefore follows immediately from the definition of $\sigmae$ (equation \ref{sigmaedef}) that
\begin{equation}
\taueR = \frac{2 \tauR}{\psi},
\end{equation}
i.e.\ that the effective molecular opacity is a factor of two larger than its free-space value because the radiation intensity at the cloud surface has half its free-space value. Similarly, the flux is $\Fstar(1)=\Fstar_0/4$ because half the sky is blocked and the radiation direction is random over the other half. Using this boundary condition to integrate the one-dimensional transfer-dissocation equation (\ref{slabeqn}) with $\sigmad=0$ and the molecular absorption rate multiplied by $\phi$ then gives 
\begin{equation}
\xm = 1-\frac{\psi}{4\tauR}.
\end{equation}
We verify that these intuitive arguments in fact give the correct leading order terms in the series expansion in Appendix \ref{largesmallappendix}.
Thus we have the limiting solution to first order in $\tauR^{-1}$ for
both $\psi<2$ and $\psi>2$. We illustrate this solution for
$\tauR\rightarrow \infty$ in Figure
\ref{planeplot}.

\subsection{Numerical Solution and Analytic Approximation}
\label{numericsol}

\begin{figure*}
\plotone{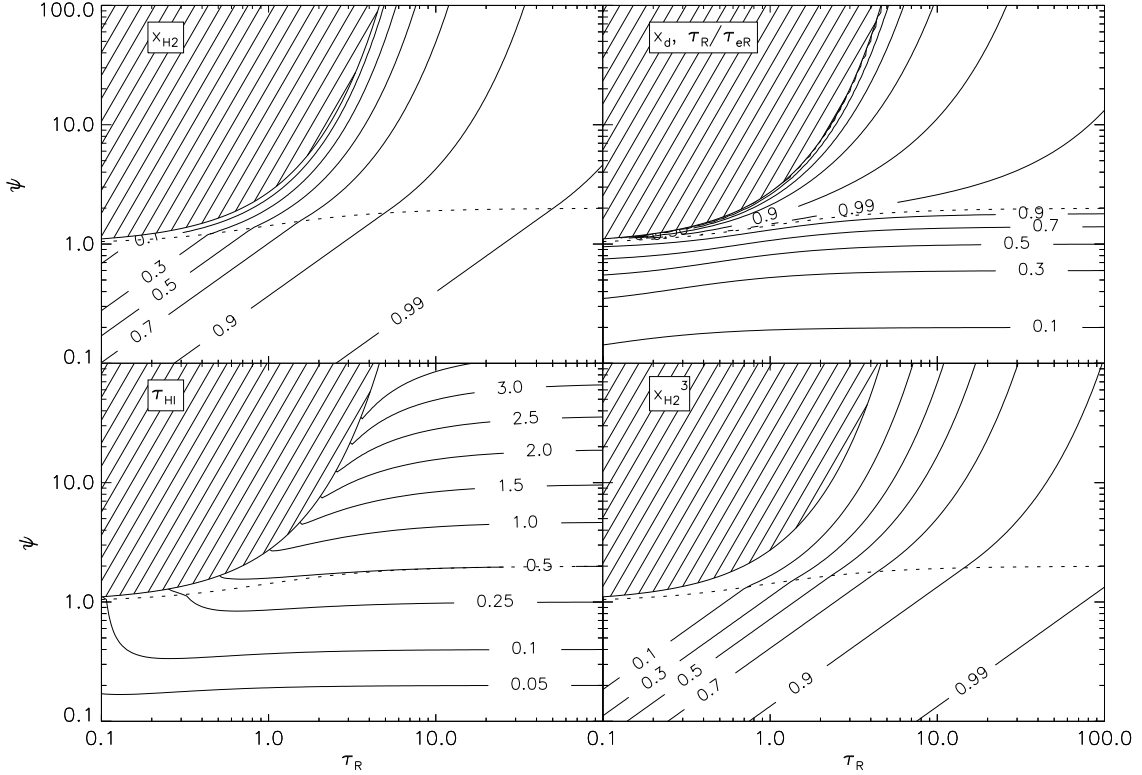}
\caption{
\label{chitaucont}
Contours showing the solution as a function of $\tauR$ and $\psi$ for
the structure of the PDR in the two-zone approximation. The values
shown are, clockwise from the upper left, $\xm$, $\xd$ or
$\tauR/\taueR$, $\tau_{\rm HI}$, and $\xm^3$. The hatched region is
the region in which there is no primarily molecular part of the cloud.
The dotted line
indicates the boundary between the presence and absence of a
dust-dominated zone. In the panel labelled $\xd$, $\tauR/\taueR$, the
contours above the dotted line indicate the value of $\xd$, those
below it show the value of $\tauR/\taueR$, and on the dotted line both
of these quantities are exactly $1.0$. We caution that the contours for $\xm = 0.1$ and $\xm=0.3$, and for $\xm^3 = 0.1$, should be regarded as giving upper limits on $\xm$, not precise estimates -- see the discussion in \S~\ref{error}.
}
\end{figure*}

We now proceed with a numerical treatment of the general case. We
solve equations (\ref{f0eqn}) and (\ref{f1eqn}), or (\ref{f0eqnodust})
and (\ref{f1eqnodust}), on a grid of points in the
$(\tauR,\psi)$-plane, and plot the results in Figure \ref{chitaucont}. In
addition to plotting $\xm$ and either $\xd$ or $\taueR$, we
also show two derived quantities of interest. The first is the dust
optical depth to the molecular transition along a radial trajectory, $\taum = nR(1-\xm) \sigmad$. We may think of this as the H\textsc{i} ``shielding column'' times the dust cross
section. The second is $\xm^3$, which is the fraction of the cloud's
volume that is within the predominantly molecular region.

The general behavior of these curves can be understood intuitively. 
If one fixes the
cloud density $n$ and dust opacity $\sigmad$, then as the cloud radius
increases so does $\tauR$, and for fixed external radiation field
$\psi$ the molecular transition moves outward, but the H\textsc{i} column to
that transition approaches a constant value. Similarly, at fixed
cloud size and hence $\tauR$, increasing the external illumination
$\psi$ raises the amount of atomic hydrogen that is required to shield
the molecules. Thus $\xm$ drops when $\psi$ increases at fixed $\tauR$.

\begin{figure}
\plotone{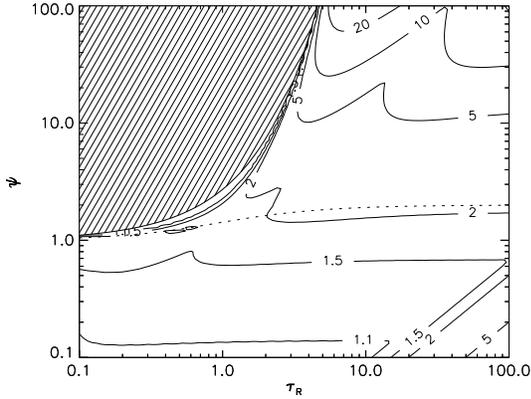}
\caption{
\label{nodust}
Contours showing the factor by which dust shielding changes the atomic or molecular volume at a given $(\tauR, \psi)$. The quantity plotted is $\mbox{max}[\xm/x_{\rm H_2,nd}, (1-x_{\rm H_2,nd})/(1-\xm)]$, where the subscript ``nd" indicates the value with no dust shielding in the limit, i.e. in the limit $\sigmad\rightarrow 0$. The quantity plotted is therefore the fractional amount by which dust shielding increases the radius of the molecular zone or decreases the radial path length through the atomic zone, whichever is larger.
}
\end{figure}

These curves also enable us to determine under what circumstances dust makes a significant contribution to shielding the gas, a subject that has been discussed considerably in the literature \citep[e.g.][]{vandishoeck86, draine96}. To evaluate the importance of dust, we can consider how the molecular and atomic volumes change as $\sigmad\rightarrow 0$. In terms of our parameters, this amounts to taking the limit as $\chi\rightarrow 0$ and $\tauR\rightarrow 0$, but the ratio $\chi/\tauR$ remains constant. Graphically, this is equivalent to sliding toward the lower right of Figure \ref{chitaucont}, along a trajectory that is close to a line of slope unity -- it is not precisely a line of slope unity because of the slight non-linearity of the relationship between $\psi$ and $\chi$. We plot the factor by which dust shielding changes the radius of the molecular zone or the radial path length through the atomic zone, whichever is larger, in Figure \ref{nodust}. As the figure shows, dust shielding changes the radius of the molecular zone by a factor of $\sim 2$ only when $\psi$ is of order unity or larger or when $\tauR$ is very large, which is about what one might expect. Dust shielding can affect the size of the molecular region even if there is no dust-dominated zone because we have allowed for dust absorptions even in the molecular shielding region. However, the effect is at most tens of percent. A larger change is possible only if the radiation field is intense enough to create a dust-dominated zone (i.e. $\psi \gtsim 1$) or the cloud is so large (i.e.\ $\tauR\gg 1$) that even weak dust shielding becomes significant because it attenuates the radiation exponentially rather than in a powerlaw fashion as do the molecules.

We can deduce approximate analytic fitting formulae for $\xm$ by
interpolating between the solutions for the limiting cases. The
following fitting formulae are reasonably accurate and can be evaluated with no numerical iteration:
\begin{equation}
\label{xmapprox}
\xm \approx
\left\{
\begin{array}{ll}
\frac{\psi^4 x_{\rm s} + (3\tauR)^4 x_{\ell}}
{\psi^4+(3\tauR)^4}, \quad &
\psi < 1 \\
\max(x_{\rm s}, x_{\ell}), \quad &
1<\psi<\psi_{\rm b} \\
\left(x_{\rm s}^{-3/2}+x_{\ell}^{-3/2}\right)^{-2/3}, \quad &
\psi_{\rm b} < \psi < 2 \\
\left(x_{\rm s}^{-3}+x_{\ell}^{-3}\right)^{-1/3}, \quad &
2 < \psi < e^{\tauR} \\
0, \quad & e^{\tauR} < \psi
\end{array}
\right.
\end{equation}
where $\psi_{\rm b} = (1.4+2\tauR)/(1.4+\tauR)$ is the approximate
value of $\psi$ at the boundary between the existence and
non-existence of a zone of dust-dominated opacity,
\begin{equation}
\label{xstapprox}
x_{\rm s}^2 =  
\left\{
\begin{array}{ll}
\left(\frac{\tauR}{\sqrt{3}\psi} + \frac{2\sqrt{3}\tauR^2}{5\psi^2}\right)^2, 
\quad & \psi < 1 \\
1-\frac{1}{4\tau_{\rm
b}^2}\left[\ln\left(1 + 2 \tau_{\rm b}-4\frac{\tau_{\rm
b}}{\psi}\right)\right]^2,
\quad & 1 < \psi < \psi_{\rm b} \\
\left[\frac{1536}{25\tauR(\tauR+2)^3}\right]^{1/2}
\left(\frac{e^{\tauR}}{\psi}-1\right)^{5/2}, 
\quad & \psi_{\rm b} < \psi
\end{array}
\right.
\end{equation}
is the approximate value of $\xm$ in the strong radiation (for $\psi>1$) or small cloud (for $\psi<1$) limits (note that we only evaluate $x_{\rm s}$ when $\psi < e^{\tauR}$), and
\begin{equation}
\label{xweapprox}
x_{\ell}^2 = 
\left\{
\begin{array}{ll}
\left(\max\left[1-\frac{\psi}{4\tauR}, 0\right]\right)^2,
\quad & \psi < \psi_{\rm b} \\
1-\frac{1}{4\tau_{\rm
b}^2}\left[\ln\left(1 + 2 \tau_{\rm b}-4\frac{\tau_{\rm
b}}{\psi}\right)\right]^2
\quad & \psi_{\rm b} < \psi < 2 \\
\left[1 - \left(\taud+\frac{\psi e^{-\taud}}{4+2\taud}\right)\frac{1}{\tauR}\right]^2,
\quad & 2 < \psi
\end{array}
\right.
\end{equation}
is the approximate value of $\xm$ in the large cloud limit.
Here
\begin{equation}
\taud={\rm E}_2^{-1}\left(\frac{2}{\psi}\right) \approx 0.83 \,\ln(0.2\psi+0.6)
\end{equation}
is the approximate optical depth
to the dust-molecular opacity crossover in the large cloud limit
when dust shielding is important,
\begin{equation}
\tau_{\rm b} \approx 1.4 \frac{\psi-1}{2-\psi}
\end{equation}
is the value of $\tauR$ at the dust-no dust
boundary for a given $\psi$, and for convenience we have used the approximation ${\rm E}_3(x) \approx e^{-x}/(2+x)$. Note that these approximations can fail
if one is very near the dust-no dust boundary because the
approximation $\tau_{\rm b}
\approx 1.4(\psi-1)/(2-\psi)$ is insufficiently accurate; in this case
one may still use the approximate expressions by replacing $\tau_{\rm
b}$ with a more accurate value of $\tauR$ on the dust-no dust boundary
computed as described in \S~\ref{dustboundary}.

\begin{figure}
\plotone{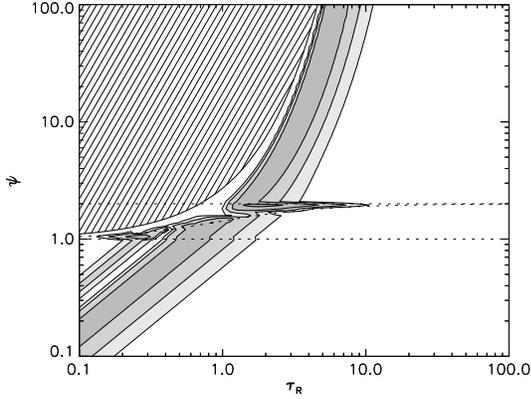}
\caption{
\label{chitaufit}
Error in the approximate analytic fit given in \S~\ref{numericsol} as
a function of $\tauR$ and $\psi$. The shaded regions indicate errors
below $2.5\%$ (\textit{no shading}), $2.5-5\%$, $5-10\%$, $10-20\%$, and $>20\%$ (\textit{darkest shading}). The maximum error is 24\%. The dotted lines show the boundaries of our
different approximation regions: $\psi<1$, $1<\psi<\psi_{\rm b}$,
$\psi_{\rm b} < \psi < 2$, and $2<\psi$. The
hatched region is $\psi>e^{\tauR}$, where there is no
predominantly molecular core. Note that the error jumps at the $\chi=1$ and $\chi=2$ lines because the fitting formula is slightly discontinuous there.
}
\end{figure}

Figure \ref{chitaufit} shows the error in our analytic approximation
as a function of $\tauR$ and $\psi$, where we define the error
as $|\xm-x_{\rm H_2,approx}| / \max(\xm,1-\xm)$ and $\xm$ is the
solution obtained by numerically solving the appropriate equations.
As the plot shows the fitting formulae are generally good to the $\sim
10\%$ level, as good as the two-zone approximation itself. The
maximum error over the range $0.1<\tauR<100$ and $0.1<\psi<100$ is
24\%, and occurs near $\tauR=1.8$, $\psi=1.9$.

We can also obtain an even simpler approximation formula if we specialize to the case where there is no or almost no dust-dominated zone and $\xm\gtsim 0.5$, which we show in Paper II is the most common case in nearby galaxies. Consider equation (\ref{f1eqnodust}), which describes the surface flux for the case of no dust. The two terms on the LHS represent the contributions to the flux from rays that do and do not pass through the cloud, respectively. If the cloud has a significant molecular core, $\xm \gtsim 0.5$, then only for a small range of angles do rays pass through the cloud but not strike the opaque molecular core, and thus the first term on the LHS is small in comparison to the second. For convenience we define
\begin{equation}
\delta = \frac{1-(1-2\mumol \taueR) e^{2\mumol\taueR}}{2\taueR^2},
\end{equation}
which enables us to rewrite equation (\ref{f1eqnodust}) as
\begin{equation}
\xm^3 = 1 - \frac{3\psi}{4\tauR}\left(1-\delta\right) \approx 1 - \frac{3\psi}{4\tauR}\left(\frac{1}{1+\delta}\right)
\end{equation}
where $\delta$ is a small, positive number. Now consider how $\delta$ varies with $\tauR$: we show in \S~\ref{limitsection} that for either small or large $\tauR$, to first order $\taueR\propto \tauR/\psi$. If we consider the series of expansion of $\delta$, this implies that $\delta$ approaches a constant at small $\tauR$ and varies as $\psi^2/\tauR^2$ for large $\tauR$. To generate our approximation we adopt an intermediate scaling
\begin{equation}
\delta \approx a \frac{\psi}{\tauR},
\end{equation}
where $a$ is a constant to be chosen to optimize the approximation. This gives
\begin{equation}
\label{chrisapprox}
\xm^3 \approx 1 - \frac{3\psi}{4(\tauR+a\psi)}.
\end{equation}
For the choice $a=0.2$, equation (\ref{chrisapprox}) agrees with the numerical solution to equations (\ref{f0eqn}) and (\ref{f1eqn}) or (\ref{f0eqnodust}) and (\ref{f1eqnodust}) to better than 15\% whenever $\psi<3$ and equation (\ref{chrisapprox}) gives $\xm^3 > 0.15$. A corresponding approximate formula for the optical depth through the atomic layer is
\begin{equation}
\label{chrisapprox1}
\taum \approx \frac{\tauR \psi}{4\tauR-a'\psi},
\end{equation}
with $a'=\frac{3}{2}-4a$. For $a=0.2$ this expression agrees with the numerical solution to better than 15\% for $\psi<3$ whenever equation (\ref{chrisapprox}) gives $\xm^3>0.1$.

\subsection{Comparison to the One-Dimensional Case}
\label{planarcomp}

\begin{figure}
\plotone{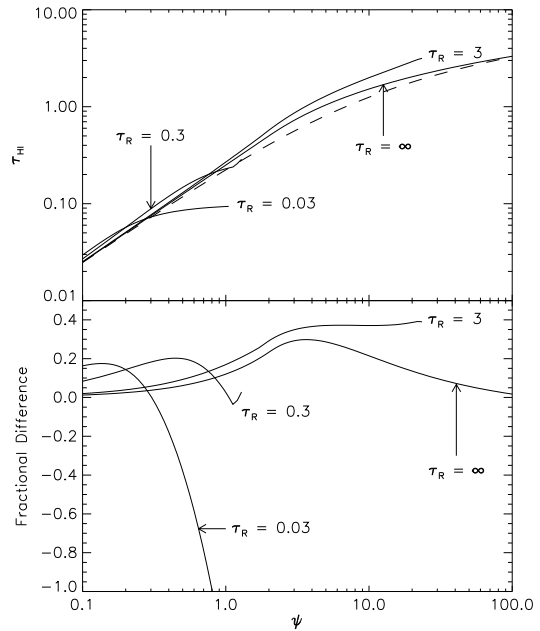}
\caption{
\label{slabcomp}
The upper panel shows the dust optical depth to the point where the
gas becomes predominantly molecular $\tau_{\rm HI}$ for an isotropic
radiation field of normalized intensity $\psi$ and various values of $\tauR$ (\textit{solid lines}) and for a unidirectional radiation field of normalized intensity $\psi/2$ (\textit{dashed line}). The lower panel shows the fractional difference between the results for finite $\tauR$ and isotropic radiation and for an infinite slab illuminated by unidirectional radiation,
defined as $\mbox{Difference} = [\tau_{\rm HI,iso}-\tau_{\rm
HI,beam}]/\tau_{\rm HI,iso}$. In
all cases the curves for finite $\tauR$ end at the value of $\psi$ for
which fully molecular region disappears ($\psi=e^{\tauR}$).
}
\end{figure}

Now that we have solved the spherical case, we are in a position to
compare to the case of a one-dimensional beam of radiation striking an infinite slab that is
often treated in the literature. This will allow us to determine when this approximation 
yields reasonably accurate results, and when it is it gives significantly different results. Figure \ref{slabcomp} shows a comparison between our
solution with isotropic radiation and varying cloud sizes versus the most common approximation in the literature: an infinite cloud and a beam of radiation whose photon number density is half the free-space value. In the calculations for finite clouds and isotropic radiation we end each curve at the value of $\psi$ for which the fully molecular region vanishes. For the beamed radiation and infinite cloud case, we use the analytic solution described in \S~\ref{slabanalyt}.

As the plot shows, when $\tauR \ll \psi$, the one-dimensional slab approximation can
produce significantly different estimates of the depth of the dust shielding layer than does our higher-dimensional approach. The difference
becomes larger as we consider smaller
clouds. For $\tauR > 1$ the slab approximation generally
underestimates the depth of the atomic layer by tens of percent,
primarily because it assumes neglects the photodissociation provided
by non-radial rays. Even for a cloud that is infinitely large, 
$\tauR=\infty$, this difference between an isotropic radiation field and a beamed one can be significant at moderate $\psi$ because even though there are no rays reaching a given position from the ``back side" of the cloud, $\mu<0$, when the radiation field is isotropic there are still non-radial rays that raise the photodissociation rate at a given position above what it would be in the purely beamed radiation field of smaller intensity.

For $\tauR < 1$ the sign of the error depends on $\psi$. When the
radiation field is weak, the slab approximation also underestimates
the depth of the atomic layer, for the same reason as when
$\tauR>1$. When the radiation field becomes strong, though, the sign
of the error reverses, although as we discus in \S~\ref{error} our fiducial model is of limited accuracy for small $\tauR$ and large $\psi$.

Physically, clouds of a wide range of sizes and densities are of
course present in the ISM. For the atomic envelopes of GMCs in the
Milky Way, a typical density is $n\sim 30$ cm$^{-3}$ and a typical
dust cross-section is $10^{-21}$ cm$^2$, so that $\sim 10$ pc of path
provides an optical depth of about 1. Since these envelopes are a few
tens of pc in size, a typical one might have $\tauR$ of a few, in
which case the slab treatment underestimates the true size of the
envelope at the tens of percent level. In low-metallicity galaxies
with low molecular fractions, however, the error is likely to be much
worse because $\tauR$ will be significantly smaller.

\subsection{Uncertainties in Spherical Geometry}
\label{error}

\begin{figure*}
\plotone{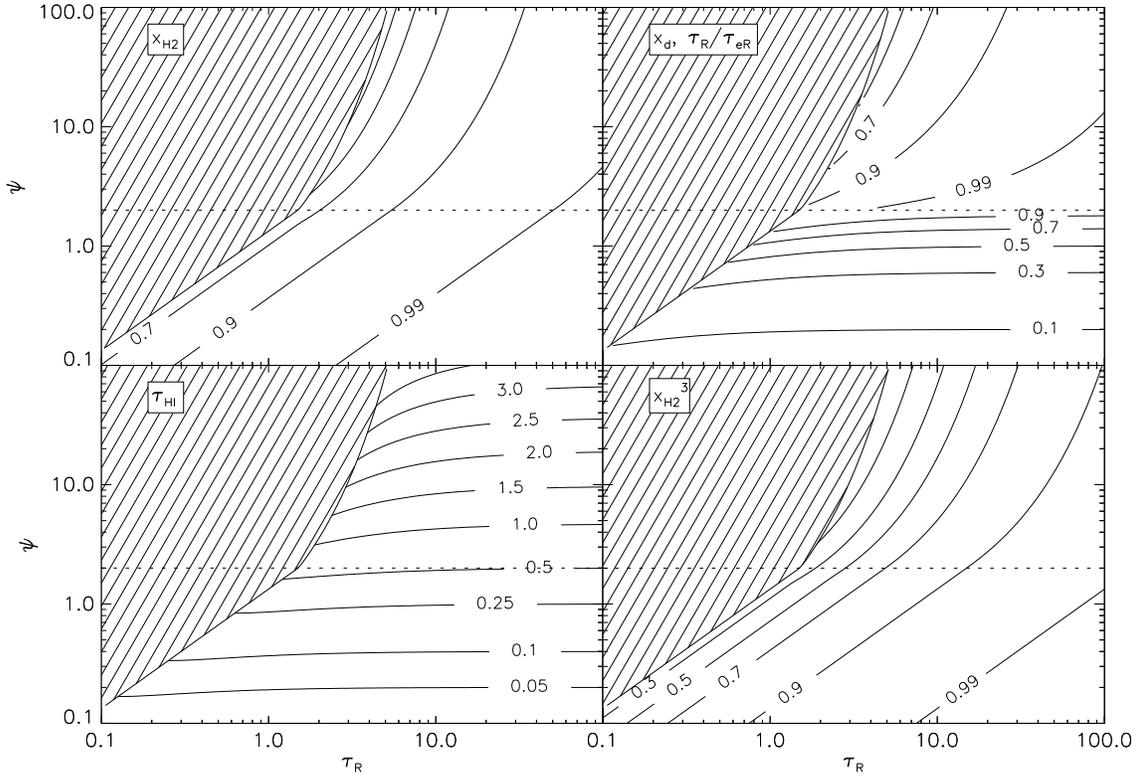}
\caption{
\label{chitaucontho}
Same is Figure \ref{chitaucont}, but assuming infinite attenuation along rays that enter the  molecular absorption-dominated region (equations \ref{f0eqnho}--\ref{nodustho}).
}
\end{figure*}

We have shown that in the case of a one-dimensional beam of radiation impinging on a slab, the two-zone approximation is capable of determining the neutral hydrogen shielding column to better than $\sim 50\%$ accuracy. This characterizes the level of error imposed by most of our physical assumptions. However, in spherical geometry we have an additional uncertainty, imposed by the fact that we must assign an effective optical depth to rays that pass at arbitrary angles through the region where molecular shielding dominates, but the gas is not yet fully molecular. In particular, rays from the ``back side" of our cloud, those with $\mu<0$, contribute to the energy density and flux throughout the cloud. Such rays are absent in the case of an infinite planar cloud, because rays with $\mu<0$ pass through the fully  molecular region and are therefore infinitely attenuated.

In this section, we seek to determine how much additional uncertainty is introduced into our calculations in spherical geometry by this complication. To do so, we note that in our treatment above we assume that the opacity in the molecular shielding region will be roughly equal to that at its surface, i.e.\ that it does not rise sharply until one is very close to the transition to fully molecular gas. The represents a minimum attenuation along $\mu<0$ rays. To check the importance of that assumption, we consider an extreme assumption in the opposite direction: that all rays with $\mu<0$ are infinitely attenuated. This assumption is obviously unphysical, since if it were true then the transition to fully molecular would occur as soon as self-shielding began to dominate over dust shielding. However, it provides a worst case with which we can compare our fiducial model as a way of characterizing our uncertainty. Since, as we shall see, the value of $\xm$ that we obtain by making this assumption is always smaller than what we obtain for the fiducial case, we may regard the fiducial case as giving upper limit on $\xm$ and the case we calculate here as giving a lower limit.

\begin{figure*}
\plotone{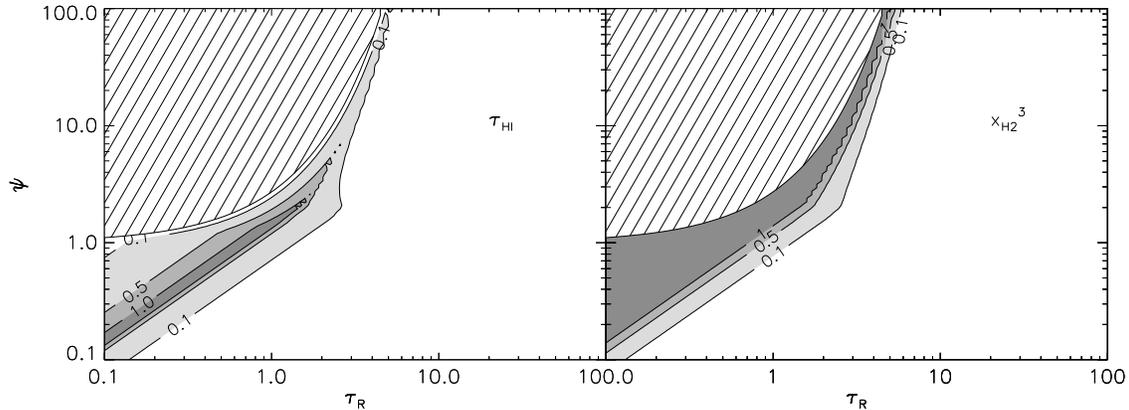}
\caption{
\label{chitaucomp}
The plots show the difference between $\taum$ and $\xm^3$ computed under the fiducial assumption that the opacity throughout the molecule-dominated region equal to that at its surface (\S~\ref{spheredust} and \S~\ref{spherenodust}) and under the assumption of infinite attenuation in this region (\S~\ref{error}).  The difference is defined as $|\tau_{\rm HI,fiducial}-\tau_{\rm HI,attenuated}|/\tau_{\rm HI,fiducial}$, and similarly for $\xm^3$. The hatched region is the region in which there is no predominantly molecular core under either assumption; the difference in this region is obviously zero.
}
\end{figure*}

If we take the intensity along rays with $\mu<0$ to be zero, this is equivalent to replacing $\eta_0(\xd,\xm;\tauR)$ with $\eta_0(\xd,\xd;\tauR)$ in the equations derived in \S~\ref{spheredust} and \S~\ref{spherenodust}, and similarly for $\eta_1$. Doing this and simplifying gives
\begin{eqnarray}
\label{f0eqnho}
\eta_0(\xd,\xd;\tauR) & = & \frac{1}{\psi} \\
\eta_1(\xd,\xd;\tauR) & = & \frac{\tauR\xd}{3\psi}\left[1-\left(\frac{\xm}{\xd}\right)^3\right]
\end{eqnarray}
for $\psi>2$ (dust shielding is significant) and
\begin{equation}
\label{nodustho}
\xm=\left(1-\frac{3\psi}{4\tauR}\right)^{1/3}
\end{equation}
for $\psi<2$ (dust shielding is not significant). Note that in this case when dust shielding is negligible it is possible to solve the equations analytically, which we have done to obtain equation (\ref{nodustho}). It is immediately obvious from this equation that the fully molecular region vanishes in the region $4\tauR/3<\psi<2$. We plot the solutions to equations (\ref{f0eqnho})--(\ref{nodustho}) in the $(\tauR,\psi)$ plane in Figure \ref{chitaucontho}, and we show the difference between this solution and our fiducial one in Figure \ref{chitaucomp}.

As the plots show, the difference between the two models is negligibly small over most of parameter space;  there is a significant difference only in the region roughly bounded by the curves $\psi \la e^{\tauR}$, $\psi\ga 4\tauR/3$, and $\psi \la 2$. Alternately, we can phrase these constraints in terms of values of $\xm^3$. This is particularly useful for $\psi < 1$, where contours of both constant $\xm$ and constant uncertainty are both straight lines corresponding to fixed $\psi/\tauR$. The 10\%, 50\%, and 100\% uncertainty contours in $\xm^3$, shown in the right panel of Figure \ref{chitaucomp}, correspond to values of $\xm^3 = 0.47$, $0.28$, and $0.20$, respectively, as computed using our fiducial model and shown in Figure \ref{chitaucomp}. Since we have already established that the two-zone approximation is uncertain at the tens of percent level, the geometric uncertainty is probably only dominant when $\psi < 1$ and our predicted molecular volume fraction $\xm^3$ is less than about a quarter. If $\psi \ga 1$, the errors at a given value of $\xm^3$ are considerably smaller, so the geometric uncertainty is not important except for clouds with very small molecular fractions. At such molecular fractions, one should interpret our fiducial case as giving only an upper limit on the molecular content of a cloud.

The significant geometric uncertainty for such clouds is not surprising, since these clouds are near the limit of having no molecules at all. For them any change in our physical assumptions that increases or reduces the amount of shielding even a small amount produces a significant change in the results. Indeed, our fiducial calculation shows some unphysical behavior in this regime, in that we find that for $\psi<1$ there is no finite value of $\tauR$ for which the fully molecular region vanishes and the cloud remains atomic throughout. This seems unlikely, given that for a chosen value of $\psi$ and very large $\tauR$ the thickness of the atomic region approaches a finite value; one would expect that clouds much smaller than this should be atomic throughout regardless of their shape, and indeed under the assumption of infinite attenuation for backside rays that we make in this section, the molecular core always vanishes at some finite value of $\tauR$ for any finite $\psi$.

Fortunately, as we discuss in \S~\ref{example}, for realistic parameters describing giant atomic-molecular complexes, we generally have $\psi\ga 1$, $\tauR\ga 1$, and, as we show in \S~\ref{example}, $\xm^3 \ga 0.5$, and in this part of parameter space the uncertainty introduced by the $\mu<0$ rays in spherical geometry is $\la 10\%$.

\subsection{Example Calculations}
\label{example}

Here we provide examples that illustrate the use of our analytic approximations for PDR structure. Since these calculations are intended to be illustrative rather than to analyze real situations (which we will discuss in the next paper in this series), we choose parameters to yield examples that span the possible combinations of parameters without worrying how well they agree with observations. As a first case, consider a ``typical" Milky Way cloud, with $\sigmad=1.1\times 10^{-21}$ cm$^2$, $n=30$ cm$^{-3}$, $\calr=3\times 10^{-17}$ cm$^3$ s$^{-1}$, and $c \Estar_0=10^8$ cm$^{-2}$ s$^{-1}$. Note that we have used a value of $\Estar_0$ somewhat larger than the solar neighborhood value because most molecular clouds are closer to the galactic center, where the radiation field is more intense. This combination of parameters gives $\chi = \fdissoc \sigmad c \Estar_0/(n\calr) = 12.2$ and $\psi=\chi (2.5+\chi)/(2.5+\chi e)=6.3$. Now consider a giant molecular cloud complex with a radius $R=50$ pc, which gives $\tauR=n\sigmad R=4.64$. Since $\psi>2$, this cloud has a significant dust shielding zone, and since $\psi < e^{\tauR}$ it also has a fully molecular core, as we expect. Using the approximation equations (\ref{xmapprox})--(\ref{xweapprox}) for this case, we find $\taud=0.52$, $x_{\ell} = 0.74$, $x_{\rm s}=14.1$, and an approximate value of $\xm\approx 0.74$. (Note that although $\xm$ is strictly less than unity, it is possible for $x_{\rm s}$ to be larger than unity because we only retained a finite number of terms in the series expansion used to generate it. However, due to the way $x_{\rm s}$ and $x_{\ell}$ are combined, our approximate expression for $\xm$ is always less than unity.) Numerical solution for these parameters gives $\xm=0.70$. Such a cloud is 34\% molecular by volume, and is shielded by an atomic column that is 15 pc deep and has a column density of $\nhi=1.4\times 10^{21}$ cm$^{-2}$ from the edge the cloud to the edge of the molecular zone.

Now consider moving this cloud to a point farther out in the Galaxy where the ambient FUV radiation field is weaker, so that all cloud parameters remain the same but now $c \Estar_0=2\times 10^7$ cm$^{-2}$ s$^{-1}$, a factor of 5 below our previous value. In this case we have the same $\tauR$, but $\chi=2.44$ and $\psi=1.43$. For this $\tauR$ we have $\psi_{\rm b}=1.40$, slightly smaller than $\psi$, so this cloud just barely still has a dust-dominated zone. Evaluating equations (\ref{xmapprox})--(\ref{xweapprox}), we have $x_{\ell}=0.87$ and $x_{\rm s} = 95$, so get an approximate value $\xm\approx  0.87$; the numerical solution is $\xm=0.92$. Thus moving the cloud to this reduced-radiation environment raises the molecular volume fraction to 77\%, and reduces the H\textsc{i} shielding column to a layer 4 pc deep containing a column of $\nhi=3.7\times 10^{20}$ cm$^{-2}$ hydrogen atoms. If the cloud were slightly denser, $n=40$ cm$^{-3}$ instead of 30 cm$^{-3}$, then $\tauR$ would increase to 6.18 from 4.64, and $\psi$ would decrease to $1.11$ from 1.43. Since $\psi_{\rm b}=1.82$ in this case, the cloud would be dominated by molecular absorption throughout. Evaluating the approximation equations gives $x_{\ell} = 0.95$, $x_{\rm s} = 0.35$, and $\xm\approx 0.95$; the numerical solution is $\xm=0.95$. Thus, the increase in density would slightly increase the molecular volume and the column density through the shielding layer to 87\% and $\nhi=5.2\times 10^{20}$ cm$^{-2}$, respectively.

Finally consider a cloud in a low-pressure dwarf galaxy with a very low star formation rate, so the cloud has lower density and metallicity than a Milky Way cloud, $n=10$ cm$^{-3}$, $\sigmad=2.2\times 10^{-22}$, and $\calr=6\times 10^{-18}$, and is exposed to a lower level of radiation, $c \Estar_0=10^6$ cm$^{-3}$ s$^{-1}$. We keep the cloud radius unchanged. This cloud has $\tauR=0.31$ and $\psi=0.29$, which from our approximate formulae gives $x_{\ell}=0.77$, $x_{\rm s}=1.4$, and $\xm\approx 0.77$. The numerical solution is $\xm=0.73$. This cloud would be 38\% molecular by volume, and would have a shielding layer of $\nhi=4.2\times 10^{20}$ cm$^{-2}$, 14 pc deep.

\section{Summary and Conclusion}
\label{conclusion}

In this paper we develop an approximate analytic solution to the
problem of determining the size of the PDR that
bounds a cloud of gas embedded in a dissociating background radiation
field. This is a reasonable approximation to the problem of finding
the location of the transition between the atomic envelope and the
molecular core in a giant atomic-molecular cloud complex, such as
those which contain the bulk of the molecular gas in the Milky Way.

We show that the location of the transition is determined by two
dimensionless parameters. These are $\tauR$, the dust optical depth
through the cloud, and $\chi$, the ratio of the rate at which
dissociating photons are absorbed by dust grains to the rate at which
they are absorbed by H$_2$ molecules in the absence of any shielding. We may intuitively
think of these parameters as characterizing the size of the cloud and
the intensity of the radiation field to which it is subjected. Within
this parameter space we identify two critical curves, which define the
boundaries at which a fully molecular region in the cloud center
appears, and at which dust shielding begins to contribute
significantly to the shielding of H$_2$ molecules. We develop the
equations that determine the sizes of the molecular and atomic
regions in this parameter space, and we provide an approximate
analytic solution for them (equations \ref{xmapprox} -- \ref{xweapprox} and equations \ref{chrisapprox} and \ref{chrisapprox1}). Our solutions are accurate to tens of percent for clouds that are $\gtsim 20\%$ molecular by volume, and provide upper limits on the molecular content at this accuracy for clouds with lower molecular content.
Using this formalism we find that for typical giant atomic-molecular complexes in the Milky Way $\chi\sim 1$, which indicates that dust shielding and self-shielding each make order unity contributions to determining the location of the atomic-molecular transition. 

Our work shows that the procedure of determing the structure of PDRs
by treating them as semi-infinite slabs illuminated by unidirectional beams of
dissociating radiation is a reasonable approximation for extremely
opaque clouds, but that it fails badly for small clouds or weak
radiation fields, i.e.\ in cases where the transition from atomic to
molecular is sufficiently far into the cloud that the cloud's
curvature cannot reasonably be neglected. In such cases the slab
approximation can either overestimate or underestimate the size of the
atomic layer by factors of order unity, depending on the particular
parameters of the cloud and the ambient radiation field.

The development of an analytic model for the structure of the atomic
envelopes of finite molecular clouds opens up the possibility of
developing a more general theory of the atomic to molecular ratio in
galaxies. In a galaxy, the mean interstellar radiation field and the
conditions in the atomic portion of the atomic interstellar medium are
determined by the star formation rate, which determines the abundance
of young, hot stars. In turn, the star formation rate depends on the
fraction of the ISM of that galaxy in molecular form, and therefore
available for star formation. At some level, therefore, star formation
in galaxies much be a self-regulating process, with the formation and
dissociation of molecular clouds representing one step in that
regulation. Developing a simple model for how the molecular fraction
in a cloud is determined by its properties and those of the ambient
radiation field represents a step toward a complete theory of the star
formation rate. In future work, we plan to develop this
theory further by applying the model demonstrated here to molecular
clouds in galaxies.

\acknowledgements We thank B.~Draine and J.~Goodman for helpful discussions, and the anonymous referee for useful comments. Support
for this work was provided by NASA through Hubble Fellowship grant
\#HSF-HF-01186 awarded by the Space Telescope Science Institute, which
is operated by the Association of Universities for Research in
Astronomy, Inc., for NASA, under contract NAS 5-26555 (MRK), and by
the National Science Foundation through grants AST-0098365 (to CFM) and PHY05-51164 (to the Kavli Institute for Theoretical Physics, where MRK, CFM, and JT collaborated on this work). JT gratefully acknowledges the support of Gilbert and Jaylee Mead for their namesake fellowship in the Yale Center for Astronomy and Astrophysics.

\appendix

\section{Evaluation of $\eta_0$ and $\eta_1$}
\label{fintegrate}

Here we evaluate the two functions
\begin{eqnarray}
\eta_0(\xd,\xm;\tauR) & = & \frac{1}{2} \int_{\mumol}^1 d\mu\, \exp(-\tauR
\xi) \\
\eta_1(\xd,\xm;\tauR) & = & \frac{1}{2} \int_{\mumol}^1 d\mu\, \mu \exp(-\tauR
\xi).
\end{eqnarray}
To evaluate the integrals, we change the variable of integration from
$\mu$ to $\xi$. Using the definition of $\xi$ (equation \ref{xidef}),
we find that
\begin{equation}
\mu = \frac{(1-\xd)(1+\xd)-\xi^2}{2\xd\xi}
\end{equation}
and
\begin{equation}
\frac{d\mu}{d\xi} = -\frac{(1-\xd)(1+\xd)+\xi^2}{2\xd\xi^2}.
\end{equation}
Making the change of variable, we find that
\begin{eqnarray}
\eta_0(\xd,\xm;\tauR) & = &
-\frac{1}{4\xd}\left[
\int_{\xim}^{1-\xd} d\xi\, \exp(-\xi \tauR) + 
(1-\xd)(1+\xd)\int_{\xim}^{1-\xd}
d\xi\, \xi^{-2} \exp(-\xi \tauR)
\right]\nonumber\\ \\
\eta_1(\xd,\xm;\tauR) & = &
\frac{1}{8\xd^2}\left[
\int_{\xim}^{1-\xd} d\xi\, \xi \exp(-\xi \tauR) -
(1-\xd)^2(1+\xd)^2
\int_{\xim}^{1-\xd} d\xi\, \xi^{-3} \exp(-\xi \tauR)
\right],\nonumber \\
\end{eqnarray}
where
\begin{equation}
\xim \equiv \xi(\xd,\mumol) = \sqrt{1-\xd^2 + \xd^2\mumol^2}-\xd\mumol.
\end{equation}
Of these integrals, the first one for $\eta_0$ may be evaluated directly,
while the first one for $\eta_1$ may be evaluated by parts. The second
integral on each line may be evaluated via the identity
\begin{eqnarray}
\int_{x_0}^{x_1} dx\, x^{-n} e^{-ax} & = &
\int_{x_0}^{\infty} dx\, x^{-n} e^{-ax} - 
\int_{x_1}^{\infty} dx\, x^{-n} e^{-ax} \\
& = & x_0^{1-n} \int_{1}^{\infty} du\, u^{-n} e^{-ax_0 u}
- x_1^{1-n} \int_{1}^{\infty} dv\, v^{-n} e^{-ax_1 v} \\
& = & x_0^{1-n} {\rm E}_n (a x_0) - x_1^{1-n} {\rm E}_n (a x_1),
\end{eqnarray}
where in the second step we made the change of variables $u=x/x_0$ and
$v=x/x_1$, and ${\rm E}_n$ is the exponential integral
function of order $n$, defined by ${\rm E}_n(x)=\int_{1}^{\infty} t^{-n}
e^{-x t} dt$. Using this identity gives
{\small
\begin{eqnarray}
\label{f0eval}
\eta_0(\xd,\xm;\tauR) & = &
\frac{1}{4\xd} 
\left\{
\frac{e^{-\yd\tauR}-e^{-\xim\tauR}}{\tauR} +
(1+\xd) \left[
{\rm E}_2 (\yd\tauR) - \frac{\yd}{\xim} {\rm E}_2(\xim\tauR)
\right]
\right\}\nonumber \\ \\
\eta_1(\xd,\xm;\tauR) & = &
\frac{1}{8\xd^2} \left\{
\frac{(1+\xim\tauR) e^{-\xim\tauR}-(1+\yd\tauR)e^{-\yd\tauR}}{\tauR^2}
+ (1+\xd)^2\left[
{\rm E}_3(\yd\tauR)-\frac{\yd^2}{\xim^2} {\rm E}_3(\xim\tauR)
\right]
\right\},
\label{f1eval}
\end{eqnarray}
}
where $\yd\equiv 1-\xd$. Note that exponential integrals obey the
recurrence relation $n{\rm E}_{n+1}(x) = e^{-x} - x {\rm E}_n(x)$, so
we could alternately have written these in terms of ${\rm E}_1(x)$ or
the classical exponential integral ${\rm Ei}(x) = - {\rm E}_1(-x)$.

\section{Solution by Series Expansion in the Strong Radiation Limit}
\label{strongappendix}

Here we solve equations (\ref{f0eqn}) and (\ref{f1eqn}) in the strong radiation limit, i.e. $\xm\ll 1$, $\xd\ll 1$, and $\mumol+1 \ll 1$, by means of series expansion. Let $\beta=\xd \tauR$ and $\gamma = 1+\mumol$. Then we have
\begin{equation}
\label{betaexpansion}
e^{-\tauR \xi} = e^{-\tauR} 
\left[1 + \mu \beta + \left(\frac{1}{2\tauR}+\frac{\tauR+1}{2\tauR}\mu^2\right) \beta^2
+ \left(\frac{1}{2\tauR} \mu + \frac{\tauR-3}{6\tauR} \mu^3\right) \beta^3 + O(\beta^4)\right].
\end{equation}
Note that we have retained terms out to order $\beta^3$. We shall see below that this is required for a consistent solution. Using this expansion in the integrals
\begin{eqnarray}
\eta_0(\xd,\xm;\tauR) & = & \frac{1}{2} \int_{\gamma-1}^{1} d\mu\, e^{-\tauR\xi} \\
\eta_1(\xd,\xm;\tauR) & = & \frac{1}{2} \int_{\gamma-1}^{1} d\mu\, \mu e^{\tauR\xi}
\end{eqnarray}
and expanding in powers of $\gamma$, we find
\begin{eqnarray}
\eta_0(\xd,\xm;\tauR) & = & \frac{e^{-\tauR}}{2} 
\left[
2-\gamma+\gamma\beta+\frac{\tauR+2}{3\tauR}\beta^2 + O(\beta^4) + O(\gamma\beta^2) + O(\gamma^2)
\right] \\
\eta_1(\xd,\xm;\tauR) & = & \frac{e^{-\tauR}}{2}
\left[
\frac{2}{3}\beta + \gamma - \gamma\beta + \frac{\tauR+2}{15}\beta^3 + O(\beta^4) + O(\gamma\beta^2) + O(\gamma^2)
\right].
\end{eqnarray}
If we similarly expand the right-hand sides of (\ref{f0eqn}) and (\ref{f1eqn}) in powers of $\beta$ and $\gamma$, we obtain the two equations
\begin{eqnarray}
\label{f0eqnstrong}
\frac{e^{-\tauR}}{2} 
\left[2-\gamma+\gamma\beta+\frac{\tauR+2}{3\tauR}\beta^2\right]
& = & \frac{1}{\psi} + O(\beta^4) + O(\gamma\beta^2) + O(\gamma^2)
\\
\label{f1eqnstrong}
\frac{e^{-\tauR}}{2} 
\left[\frac{2}{3}\beta + \gamma - \gamma\beta + \frac{\tauR+2}{15}\beta^3\right]
& = & \frac{\beta}{3\psi}  + O(\beta^4) + O(\gamma\beta^2) + O(\gamma^2).
\end{eqnarray}
If we combine these two equations by eliminating the common factor $e^{\tauR}/\psi$, we obtain an equation for the relationship between $\beta$ and $\gamma$:
\begin{equation}
-\frac{\gamma}{2}+\frac{\gamma\beta}{2} + \frac{\tauR+2}{6\tauR} \beta^2 =
\frac{3}{2}\left(\frac{\gamma}{\beta}\right) - \frac{3}{2}\gamma + \frac{\tauR+2}{10\tauR} \beta^2
+ O(\beta^3) + O(\gamma\beta) + O\left(\frac{\gamma^2}{\beta}\right).
\end{equation}
The only way for this equation to have a consistent solution in which the orders on both sides balance is if $\gamma$ is of order $\beta^3$. In this case the leading order on both sides is $\beta^2$ (an order we retained only be performing the expansion in equation \ref{betaexpansion} to order $\beta^3$), and balancing the leading order terms gives
\begin{equation}
\label{gammasol}
\gamma = \frac{2\tauR+4}{45\tauR} \beta^3.
\end{equation}
Since we now know the order of all terms, we can solve equation (\ref{f0eqnstrong}) to leading order to obtain
\begin{equation}
\label{xdsol}
\xd = \frac{\beta}{\tauR} = \left[\frac{6}{\tauR(\tauR+2)}\left(\frac{e^{\tauR}}{\psi}-1\right)\right]^{1/2}.
\end{equation}
Similarly, we know that
\begin{equation}
\left(\frac{\xm}{\xd}\right)^2 = 1-\mumol^2 = 2\gamma + O(\gamma^2).
\end{equation}
Substituting (\ref{gammasol}) for $\gamma$ and (\ref{xdsol}) for $\xd$ and re-arranging, we obtain to leading order
\begin{equation}
\xm = \left[\frac{1536}{25\tauR(\tauR+2)^3}\right]^{1/4} \left(\frac{e^{\tauR}}{\psi}-1\right)^{5/4}.
\end{equation}

\section{Solution by Series Expansion in the Large and Small Cloud Limits}
\label{largesmallappendix}

Here we solve equations (\ref{f0eqnodust}) and (\ref{f1eqnodust}) by
series expansion in the limits $\tauR\rightarrow 0$ and
$\tauR\rightarrow \infty$. We approach this
problem by defining $\beta = \taueR/\tauR$, so that with some
rearrangement the equations are
\begin{eqnarray}
\label{f0ndlim}
1-e^{2\beta\mumol\tauR}+2\beta\tauR-\frac{4}{\psi}\tauR & = & 0\\
\label{f1ndlim}
(1-2\beta\mumol\tauR)e^{2\beta\mumol\tauR}-1+2\beta^2\tauR^2 -
\frac{8}{3\psi} \beta^2\tauR^3 \left[1-(1-\mumol^2)^{3/2}\right] & = &
0.
\end{eqnarray}
For the case $\tauR\rightarrow 0$ we then let
\begin{eqnarray}
\beta & = & \beta_0 + \beta_1 \tauR + \beta_2\tauR^2 + \cdots \\
\mumol & = & \mu_0 + \mu_1 \tauR + \mu_2\tauR^2 + \cdots.
\end{eqnarray}
Expanding equations (\ref{f0ndlim}) and (\ref{f1ndlim}) to
leading order in $\tauR$ and re-arranging gives
\begin{eqnarray}
2\beta_0(1-\mu_0) - \frac{4}{\psi} & = & 0 \\
2\beta_0^2(1 - \mu_0^2) & = & 0,
\end{eqnarray}
which has the solution $\mu_0=-1$, $\beta_0=1/\psi$. Using these
values and continuing the expansion to the next order, we obtain
\begin{eqnarray}
-2\beta_1 + \frac{1+\mu_1\psi}{\psi^2} & = & 0 \\
\frac{4\mu_1}{\psi^2} & = & 0,
\end{eqnarray}
so $\mu_1=0$ and $\beta_1=1/(2\psi^2)$. Continuing to one more order,
we have
\begin{eqnarray}
4\beta_2-\frac{2+6\beta_2\psi^2}{3\psi^3} & = & 0\\
-1 + 6\mu_2 \psi^2 & = & 0,
\end{eqnarray}
so $\mu_2=1/(6\psi^2)$ and $\beta_2=1/(4\psi^3)$. It improves the
accuracy of the approximation for $\xm$ significantly at small $\psi$
to include one more order, so we do so:
\begin{eqnarray}
4\beta_3+\frac{1}{3\psi^4}-\frac{2}{\psi}\mu_3 & = & 0\\
\frac{4}{\psi^2}\mu_3-\frac{8}{5\psi^5} & = & 0,
\end{eqnarray}
so $\mu_3=2/(5\psi^3)$ and $\beta_3=7/(60\psi^4)$. Therefore to order
$\tauR^4$ we have
\begin{eqnarray}
\frac{\taueR}{\tauR} & = & \frac{1}{\psi} + \frac{\tauR}{2\psi^2} +
\frac{\tauR^2}{4\psi^3} + \frac{7\tauR^3}{60\psi^4}\\
\mumol & = & -1 + \frac{\tauR^2}{6\psi^2}+\frac{2\tauR^3}{5\psi^3}\\
\xm & = & \sqrt{1-\mumol^2} =
\frac{\tauR}{\sqrt{3}\psi}+\frac{2\sqrt{3} \tauR^2}{5\psi^2}.
\end{eqnarray}

For the case $\tauR\rightarrow\infty$, we let
\begin{eqnarray}
\beta & = & \beta_0 + \beta_{-1/2}\tauR^{-1/2} + \beta_{-1}\tauR^{-1}+\cdots
\\
\mumol & = & \mu_0 + \mu_{-1/2}\tauR^{-1/2} + \mu_{-1}\tauR^{-1}+\cdots,
\end{eqnarray}
and if we expand equations (\ref{f0ndlim}) and (\ref{f1ndlim}) in
powers of $\tauR^{-1}$ then the leading order equations are
\begin{eqnarray}
2\beta_0 - \frac{4}{\psi} & = & 0 \\
-\frac{8}{3\psi} \left[1-(1-\mu_0^2)^{3/2}\right] & = & 0.
\end{eqnarray}
Therefore $\beta_0=2/\psi$ and $\mu_0=0$. Continuing the expansion to
the next order,
\begin{eqnarray}
2\beta_{-1/2} & = & 0\\
-\frac{16}{\psi^3}\mu_{-1/2}^2+\frac{8}{\psi^2} & = & 0,
\end{eqnarray}
so $\beta_{-1/2}=0$ and $\mu_{-1/2} = -\sqrt{\psi/2}$. Continuing one
more order,
\begin{eqnarray}
1+2\beta_{-1} & = & 0\\
\frac{16\sqrt{2}}{\psi^5/2}\mu_{-1} & = & 0,
\end{eqnarray}
so $\beta_{-1}=-1/2$ and $\mu_{-1}=0$
Thus to order $\tauR^{-1}$ in the limit $\tauR\rightarrow\infty$ we have
\begin{eqnarray}
\frac{\taueR}{\tauR} & = & \frac{2}{\psi} \\
\mumol & = & -\sqrt{\frac{\psi}{2\tauR}} \\
\xm & = & 1 - \frac{\psi}{4\tauR}.
\end{eqnarray}


\begin{thebibliography}{19}
\expandafter\ifx\csname natexlab\endcsname\relax\def\natexlab#1{#1}\fi

\bibitem[{{Allen} {et~al.}(2004){Allen}, {Heaton}, \& {Kaufman}}]{allen04}
{Allen}, R.~J., {Heaton}, H.~I., \& {Kaufman}, M.~J. 2004, \apj, 608, 314

\bibitem[{{Black} \& {van Dishoeck}(1987)}]{black87}
{Black}, J.~H. \& {van Dishoeck}, E.~F. 1987, \apj, 322, 412

\bibitem[{{Blitz} \& {Rosolowsky}(2006)}]{blitz06b}
{Blitz}, L. \& {Rosolowsky}, E. 2006, \apj, 650, 933

\bibitem[{{Browning} {et~al.}(2003){Browning}, {Tumlinson}, \&
  {Shull}}]{browning03}
{Browning}, M.~K., {Tumlinson}, J., \& {Shull}, J.~M. 2003, \apj, 582, 810

\bibitem[{{Cardelli} {et~al.}(1989){Cardelli}, {Clayton}, \&
  {Mathis}}]{cardelli89a}
{Cardelli}, J.~A., {Clayton}, G.~C., \& {Mathis}, J.~S. 1989, \apj, 345, 245

\bibitem[{{Draine}(1978)}]{draine78}
{Draine}, B.~T. 1978, \apjs, 36, 595

\bibitem[{{Draine} \& {Bertoldi}(1996)}]{draine96}
{Draine}, B.~T. \& {Bertoldi}, F. 1996, \apj, 468, 269

\bibitem[{{Elmegreen}(1993)}]{elmegreen93}
{Elmegreen}, B.~G. 1993, \apj, 411, 170

\bibitem[{{Elmegreen} \& {Elmegreen}(1987)}]{elmegreen87}
{Elmegreen}, B.~G. \& {Elmegreen}, D.~M. 1987, \apj, 320, 182

\bibitem[{{Federman} {et~al.}(1979){Federman}, {Glassgold}, \&
  {Kwan}}]{federman79}
{Federman}, S.~R., {Glassgold}, A.~E., \& {Kwan}, J. 1979, \apj, 227, 466

\bibitem[{{Hollenbach} \& {Tielens}(1999)}]{hollenbach99}
{Hollenbach}, D.~J. \& {Tielens}, A.~G.~G.~M. 1999, Reviews of Modern Physics,
  71, 173

\bibitem[{{Liszt}(2002)}]{liszt02}
{Liszt}, H. 2002, \aap, 389, 393

\bibitem[{{Liszt} \& {Lucas}(2000)}]{liszt00}
{Liszt}, H. \& {Lucas}, R. 2000, \aap, 355, 333

\bibitem[{{Neufeld} \& {Spaans}(1996)}]{neufeld96}
{Neufeld}, D.~A. \& {Spaans}, M. 1996, \apj, 473, 894

\bibitem[{{Roberge} {et~al.}(1981){Roberge}, {Dalgarno}, \&
  {Flannery}}]{roberge81}
{Roberge}, W.~G., {Dalgarno}, A., \& {Flannery}, B.~P. 1981, \apj, 243, 817

\bibitem[{{Spaans} \& {Neufeld}(1997)}]{spaans97}
{Spaans}, M. \& {Neufeld}, D.~A. 1997, \apj, 484, 785

\bibitem[{{Sternberg}(1988)}]{sternberg88}
{Sternberg}, A. 1988, \apj, 332, 400

\bibitem[{{van Dishoeck} \& {Black}(1986)}]{vandishoeck86}
{van Dishoeck}, E.~F. \& {Black}, J.~H. 1986, \apjs, 62, 109

\bibitem[{{Wolfire} {et~al.}(2008){Wolfire}, {Tielens}, {Hollenbach}, \&
  {Kaufman}}]{wolfire08}
{Wolfire}, M.~G., {Tielens}, A.~G.~G.~M., {Hollenbach}, D., \& {Kaufman}, M.~J.
  2008, \apj

\end{thebibliography}
\end{document}